\documentclass[reprint,
superscriptaddress,
aps,
pra,
onecolumn,
]{revtex4-1}

\usepackage{dcolumn}
\usepackage{amsmath,amssymb,graphicx,comment,bm,color,amsfonts}
\usepackage[pagebackref]{hyperref}
\usepackage{epsfig,epstopdf}
\usepackage{bm}
\newcommand{\vect}[1]{\boldsymbol{\mathbf{#1}}}

\begin{document}

\title{Coarse-grained pressure dynamics in superfluid turbulence}
\author{Jason Laurie}
\email{j.laurie@aston.ac.uk}
\affiliation{Mathematics Group, School of Engineering and Applied Science, Aston University, Aston Triangle, Birmingham, B4 7ET, United Kingdom}

\author{Andrew W. Baggaley} 
\email{andrew.baggaley@ncl.ac.uk}
\affiliation{Joint Quantum Centre (JQC) Durham--Newcastle, School of Mathematics and Statistics, Newcastle University, Newcastle upon Tyne, NE1 7RU, United Kingdom}


\date{\today}

\begin{abstract}
Quantum mechanics places significant restrictions on the hydrodynamics of superfluid flows. Despite this it has been observed that turbulence in superfluids can, in a statistical sense, share many of the properties of its classical brethren; coherent bundles of superfluid vortices are often invoked as an important feature leading to this quasi-classical behavior. A recent experimental study [E. Rusaouen, B. Rousset, and P.-E. Roche, EPL, {\bf 118}, 1, 14005, (2017)] inferred the presence of these bundles through intermittency in the pressure field, however  direct visualization of the quantized vortices to corroborate this finding was not possible. In this work, we performed detailed numerical simulations of superfluid turbulence at the level of individual quantized vortices through the vortex filament model. Through course-graining of the turbulent fields, we find compelling evidence supporting these conclusions at low temperature. Moreover, elementary simulations of an isolated bundle show that the number of vortices inside a bundle can be directly inferred from the magnitude of the pressure dip, with good theoretical agreement derived from the HVBK equations. Full simulations of superfluid turbulence show strong spatial correlations between course-grained vorticity and low pressure regions, with intermittent vortex bundles appearing as deviations from the underlying Maxwellian (vorticity) and Gaussian (pressure) distributions. Finally, simulations of a decaying random tangle in an ultra-quantum regime show a unique fingerprint in the evolution of the pressure distribution, which we argue can be fully understood using the HVBK framework.
\end{abstract}

\maketitle

\section{Introduction}
Classical turbulent flows are characterized by the chaotic motion of both the velocity and pressure fields. This motion can be thought of as an amalgamation of swirling eddies quantified via the fluid vorticity field (curl of the velocity). These eddies populate a hierarchy of scales, dictating an energy distribution across scales, and can give rise to spatial organization of the flow through the appearance of coherent structures. For classical flows, described by the Navier-Stokes equations, these coherent structures appear as vortex worms -- allantoid-shaped regions of intense vorticity. Understanding how these coherent structures and the corresponding energy density distribution appear and evolve, particularly in the limit of increasing Reynolds number, is of paramount importance to fluid dynamicists.

Increased interest in studying classical turbulence in liquid helium-4 has arisen due to the recent development of large-scale experimental setups that permit the study of turbulence at extreme Reynolds numbers, due to the lower viscosity of helium-4. This allows for the study of parameter regimes far beyond what is capable in similar sized experiments in water or air~\cite{roche_triggering_2010,rousset_superfluid_2014}. Below $T_\lambda\sim 2.17~\rm K$ helium-4 becomes superfluid and the hydrodynamic properties of the fluid become more complex. At temperatures below $T_\lambda$, helium-4 can be described by a two-fluid model~\cite{tisza_theory_1947,landau_theory_1941,landau_theory_1949}: a viscous normal fluid coexisting with an inviscid superfluid, which are coupled through the vorticity of the superfluid component. In the superfluid component, quantum mechanical constraints lead to confinement of the  superfluid vorticity to atomically thin and identical topological line defects in the superfluid density field with fixed circulation, named quantum vortices. The circulation around such vortices is quantized in units of Planck's constant over the mass of the helium-4 atom, known as the quantum of circulation $\kappa=nh/m$ for $n\in\mathbb{N}$. Superfluid turbulence is then defined as a tangle of quantized vortex lines that generate a complex, irrotational, velocity field with a corresponding singular vorticity field along the centerline of the vortices.

Due to the quantization of vorticity, an individual vortex line cannot be stretched; its length can change but the circulation is fixed. This is in contrast to classical vortices. Indeed, the stretching of classical vortices in turbulent flows is believed to be an important mechanism for energy transfer in classical turbulence~\cite{goto_physical_2008}. However, despite the absence of vortex stretching in superfluid turbulence, it has been shown to share many of the statistical properties of its classical cousin~\cite{salort_energy_2012}. It is commonly hypothesized that on a coarse-grained level, localized bundles of quantized vortex filaments lead to a macroscopic fluid flow around the bundle with a circulation equal to the sum of individual circulations. This is then interpreted as a larger scale coherent flow similar to a large-scale vortex observed in classical turbulence. The formation of bundles allows for the direct transfer of energy through the scales, as the relative motion of vortices within a bundle can mimic the stretching of a classical vortex tube. It is also expected to have a decisive impact on the dissipation of the turbulent kinetic energy. 

In the absence of structure, i.e. a random orientated tangle of vortices, vortex reconnections are believed to play the dominant role in dissipation of energy at small scales. On the other hand, if the turbulence is polarized, due to the presence of many bundles, then vortex reconnections may become suppressed and Kelvin waves, generated by reconnections or the motion of eddies at the inter-vortex scale, can cascade and become the dominant dissipation mechanism~\cite{laurie_interaction_2010,baggaley_kelvin-wave_2014}. For more information about the dynamics of individual vortex reconnections we refer the reader to a recent paper~\cite{galantucci_crossover_2019} and references contained within.

Unfortunately, experimental studies of superfluid turbulence in helium-4, are performed in cryostats at temperatures below $2\mathrm{K}$ which significantly hampers the ability to visualize or indeed measure important statistics of the flow. Hot wire probes are problematic as they cause a local heating of the superfluid leading to artificial superflow and/or transformation of the superfluid back into a normal fluid. The most common technique is to measure pressure fluctuations and to then infer the dynamics of the velocity field~\cite{maurer_local_1998,salort_turbulent_2010,salort_energy_2012,rousset_superfluid_2014}. Inspired by the group of Roche~\cite{rusaouen_detection_2017}, who recently studied the statistics of pressure fluctuations as a route for the detection of coherent structures, we perform a series of numerical simulations of superfluid turbulence in the low-temperature limit. Our aim is to investigate the dynamics and statistics of coarse-grained superfluid fields and examine the relationship between the pressure and velocity in a variety of turbulent regimes. This is crucially important, as while significant advances in the visualization of quantum turbulence have been made in the last decade~\cite{fisher_andreev_2014,guo_visualization_2014}, we are yet to have a direct visualization of coherent structures in quantum turbulence. Our numerical results provide strong support to the implied observation of Rusaouen {\it et al.}~\cite{rusaouen_detection_2017} that large-scale (larger than the mean inter-vortex spacing) vortex bundles can be inferred from strong negative pressure fluctuations in low temperature superfluid turbulence.

The structure of this article is as follows. We begin by overviewing the two main finite-temperature superfluid models - the HVBK equations and the vortex filament method, and outline the mathematical relationship between the velocity and pressure. Then, we examine how numerical coarse-graining of the superfluid velocity field can affect numerical data statistics and draw direction relation to experimental probe sizes. From this, we explore the velocity-pressure relationship in a static superfluid tangle with classical turbulence statistics before moving on to study the pressure-velocity evolution dynamics across three categories of turbulent tangles: (i) random or ultra-quantum turbulence with no apparent structure, (ii) an imposed large-scale Taylor-Green tangle, and finally (iii) a quasi-classical tangle with a Kolmogorov energy spectrum.

\section{Superfluid turbulence models}
Finite-temperature superfluids can be phenomenologically described by the two-fluid model~\cite{tisza_theory_1947,landau_theory_1941,landau_theory_1949}, which consists of an intimate mixture of an inviscid superfluid and a classical viscous normal fluid. Each fluid is associated with separate velocity and density fields, denoted ${\bf v}_n$ and $\rho_n$ for the normal fluid and ${\bf v}_s$ and $\rho_s$ for the superfluid respectively, with total fluid density $\rho=\rho_{n}+\rho_{s}$ whose ratio is strongly temperature dependent, with $\rho_{n}/\rho_{s}\to 0$ in the zero-temperature limit. The HVBK equations~\cite{hall_rotation_1956,bekarevich_phenomenological_1961} give a two-fluid description with the normal fluid modelled by the Navier-Stokes equations coupled via a mutual friction term $\mathbf{F}_{\rm mf}$ to the coarse-grained inviscid superfluid component $\mathbf{v}_{s}$ modelled by the Euler equations
\begin{subequations}
\begin{align}
\frac{\partial\mathbf{v}_{n}}{\partial t}+\left(\mathbf{v}_{n}\cdot\nabla\right)\mathbf{v}_{n}= & -\frac{1}{\rho}\nabla P+\nu_n \nabla^{2}\mathbf{v}_{n}+\frac{\rho_{s}}{\rho}\mathbf{F}_{\rm mf}, \quad \nabla\cdot {\bf v}_n=0,\label{eq:navier}\\
\frac{\partial\mathbf{v}_{s}}{\partial t}+\left(\mathbf{v}_{s}\cdot\nabla\right)\mathbf{v}_{s}= & -\frac{1}{\rho}\nabla P-\frac{\rho_{n}}{\rho}\mathbf{F}_{\rm mf}, \quad \nabla\cdot {\bf v}_s=0. \label{eq:euler}
\end{align}
\end{subequations}
Here, $P$ is the pressure, and $\nu_n$ is the kinematic viscosity of the normal fluid component. The mutual friction term ${\bf F}_{\rm mf}$ provides coupling between the normal and superfluid components and acts principally at the regions of high superfluid vorticity. The exact expression for ${\bf F}_{\rm mf}$ in the HVBK equations can be found in \cite{lvov_energy_2006}, here it is sufficient to comment that an approximate mutual friction~\cite{lvov_energy_2006} can be defined by
\begin{equation} \label{eq:F_mf}
{\bf F}_{\rm mf} \simeq \alpha \rho_s   \langle |\vect{\omega}_{s}| \rangle (\mathbf{v}_{s}-\mathbf{v}_{n}),
\end{equation}
where $\langle |\vect{\omega}_{s}| \rangle$ introduces the notion of a course-grained superfluid vorticity.

By taking the divergence of the HVBK equations, and assuming incompressibility of the fluid flow, one can relate the pressure $P$ to the vorticity through a Poisson equation involving  the spin tensors $\mathbf{W}_{i}$  and the strain tensors $\mathbf{E}_{i}$ for the superfluid $i=s$ and normal fluid $i=n$ components respectively
\begin{align}\label{eq:pressure_full}
\nabla^{2}P= & \frac{\rho_{s}}{2}\left(\mathbf{W}_{s}:\mathbf{W}_{s}-\mathbf{E}_{s}:\mathbf{E}_{s}\right)+\frac{\rho_{n}}{2}\left(\mathbf{W}_{n}:\mathbf{W}_{n}-\mathbf{E}_{n}:\mathbf{E}_{n}\right),
\end{align}

where 
\begin{align*}
\mathbf{W}_{i}= & \frac{1}{2}\left[\nabla\mathbf{v}_{i}-\nabla\mathbf{v}_{i}^{T}\right],\quad\mathbf{E}_{i}= \frac{1}{2}\left[\nabla\mathbf{v}_{i}+\nabla\mathbf{v}_{i}^{T}\right],
\end{align*}
and the operation $:$ denotes the double dot product between two matrices defined as $A:B=a_{ij}b_{ij}$ with index summation implied for matrix elements $a_{ij}$ and $b_{ij}$.

In the low-temperature limit, the superfluid density dominates over the normal fluid density, and hence the Laplacian of the pressure can be approximated by
\begin{align}\label{eq:pressure}
\nabla^{2}P & \simeq\frac{\rho_{s}}{2}\left(\mathbf{W}_{s}:\mathbf{W}_{s}-\mathbf{E}_{s}:\mathbf{E}_{s}\right).
\end{align}
Equation~\eqref{eq:pressure} indicates that in high vorticity regions where ${\bf W}_s$ is large, such as those inside coherent vortex bundles, the pressure field, $P$ will become strongly negative as can be determined if one inverts the Laplacian operator. This hints at the inherent connection between pressure and vorticity that can be used to probe the fluid structure in experimental studies.

An alternative model is to describe the  the superfluid velocity field $\mathbf{v}_{s}$ through dynamics of one-dimensional vortex lines using the vortex filament method (VFM) of Schwarz~\cite{schwarz_three-dimensional_1985,schwarz_three-dimensional_1988} which is subsequently coupled to the Navier-Stokes equations~\eqref{eq:navier} for the normal fluid component through a redefined mutual friction term. The superfluid velocity field ${\bf v}_s$ in the vortex filament model is determined by integrating the Biot-Savart law over the vortex filament tangle. The advantage of this method is that it permits a description of the superfluid velocity at scales far below the inter-vortex spacing (unlike the HVBK equations) leading to a non-coarse-grained superfluid velocity field. The vortex filament model replaces Eq.~\eqref{eq:euler} with an evolution equation for the vortex filaments of the form
\begin{align}
\frac{{\rm d}{\bf s}}{{\rm d}t} & ={\bf v}_{s}+\alpha{\bf s}'\times({\bf v}_{n}-{\bf v}_{s})-\alpha'{\bf s}'\times\left[{\bf s}'\times\left({\bf v}_{n}-{\bf v}_{s}\right)\right],\label{eq:Schwarz-1}
\end{align}
where ${\bf s}(\xi,t)$ is the position of the one-dimensional space curves representing quantized vortex filaments. Here, $\alpha$ and $\alpha'$ are the non-dimensional temperature dependent friction coefficients (for the explicit mutual friction term), ${\bf s}'={\rm d}{\bf s}/{\rm d}\xi$ is the unit tangent vector at the point $\mathbf{s}$, $\xi$ is arc length, and ${\bf v}_{n}$ is the normal fluid velocity at the point ${\bf s}$.

The friction coefficients $\alpha$ and $\alpha'$, play a crucial role in dictating the nature of the superfluid hydrodynamics. Indeed, they can be used to define a single dimensionless parameter akin to a superfluids Reynolds number~\cite{finne_intrinsic_2003},
\begin{align*}
{\rm Re}_\alpha=\frac{1-\alpha'}{\alpha}.
\end{align*}
In analogy to classical turbulence, we can expect the superfluid flow to be turbulent when ${\rm Re}_\alpha \gg 1$, which is the case for both the experiments of~\cite{rusaouen_detection_2017}, and the simulations we shall perform.

The velocity of the superfluid component ${\bf v}_{s}$ can be decomposed into a self-induced velocity generated by the vortex tangle ${\bf v}_{s}^{{\rm si}}$, and an external superfluid flow ${\bf v}_{s}^{{\rm ext}}$ such that ${\bf v}_{s}={\bf v}_{s}^{{\rm si}}+{\bf v}_{s}^{{\rm ext}}$. The self-induced velocity ${\bf v}_{s}^{{\rm si}}$ of the vortex line at the point ${\bf s}$, is computed using the Biot-Savart law~\cite{saffman_vortex_1992}
\begin{align}\label{eq:BS-1}
{\bf v}_{s}^{{\rm si}}({\bf s},t) & =\frac{\Gamma}{4\pi}\oint_{\cal L}\frac{({\bf r}-{\bf s})}{\left|{\bf r}-{\bf s}\right|^{3}}\times{\rm {\bf d}}{\bf r},
\end{align}
where $\Gamma=9.97\times10^{-4}~{\rm cm^{2}/s}$ (in $^{4}$He) and the line integral extends over the entire vortex configuration $\mathcal{L}$. The external superfluid flow ${\bf v}_{s}^{\rm ext}$ is an externally imposed irrotational flow arising through either an excitation mechanism of the superfluid component or through the conservation of total mass of helium-4 in the presence of a mean normal fluid flow.

\section{Setup}

We perform numerical simulations using the vortex filament method to study the pressure field dynamics in the low-temperature limit, where the turbulence is solely governed by the superfluid flow with the normal fluid component absent. We use the model of Eq.~\eqref{eq:Schwarz-1} with no external or normal fluid flow ${\bf v}_{s}^{\rm ext}={\bf v}_{n}=0$, but include a small mutual friction component ($\alpha=0.01$ and $\alpha'=0.0$) which models a superfluid flow at low temperature $T\approx 1.1{\rm K}$~\cite{schwarz_three-dimensional_1985}, corresponding to 99\% pure superfluid. Although our goal is to investigate zero-temperature superfluid turbulence, we add a small mutual friction to ensure that we reduce the effects of artificial numerical dissipation by our numerical scheme.

Our calculations are performed in a periodic cube of size $D=0.1~\rm cm$. The numerical technique to which vortex lines are discretized into a number of points ${\bf s}_j$ for $j=1, \cdots N$ held at a minimum separation $\Delta\xi/2$, compute the time evolution, de-singularize the Biot-Savart integrals, evaluate ${\bf v}_s$, and algorithmically perform vortex reconnections when vortex lines come sufficiently close to each other, are described in detail in previous papers~\cite{baggaley_tree_2012,baggaley_sensitivity_2012}. The Biot-Savart integral is computed using the a tree-algorithm approximation~\cite{baggaley_tree_2012} with opening angle set to $\theta=0.2$. We take $\Delta \xi=2.5\times 10^{-3}~\rm cm$ and a time step of $5\times 10^{-5}~\rm s$.

\section{A Vortex Bundle}

\begin{figure}[htp!]
\begin{center}
\includegraphics[width=0.4\columnwidth]{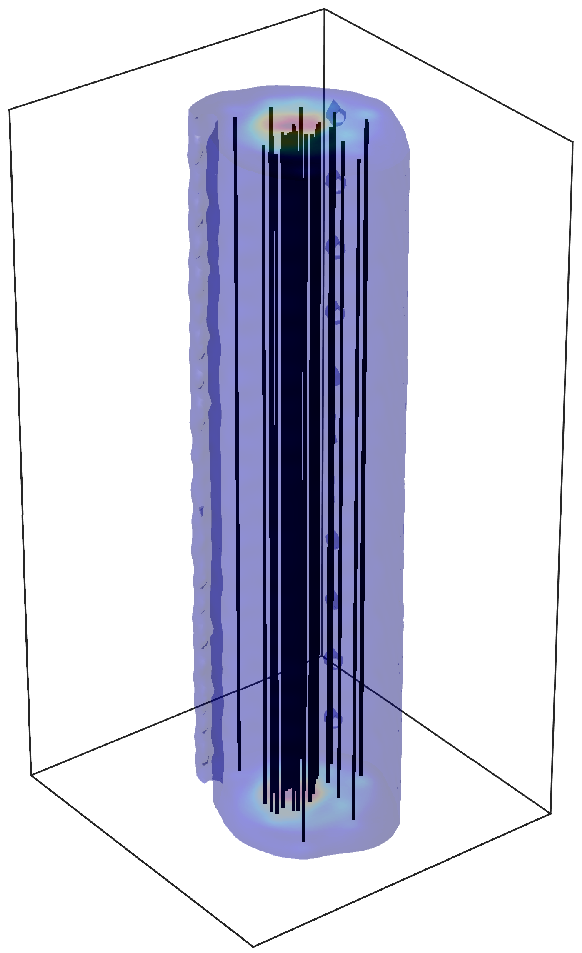}
\hspace{0.05\columnwidth}
\includegraphics[width=0.4\columnwidth]{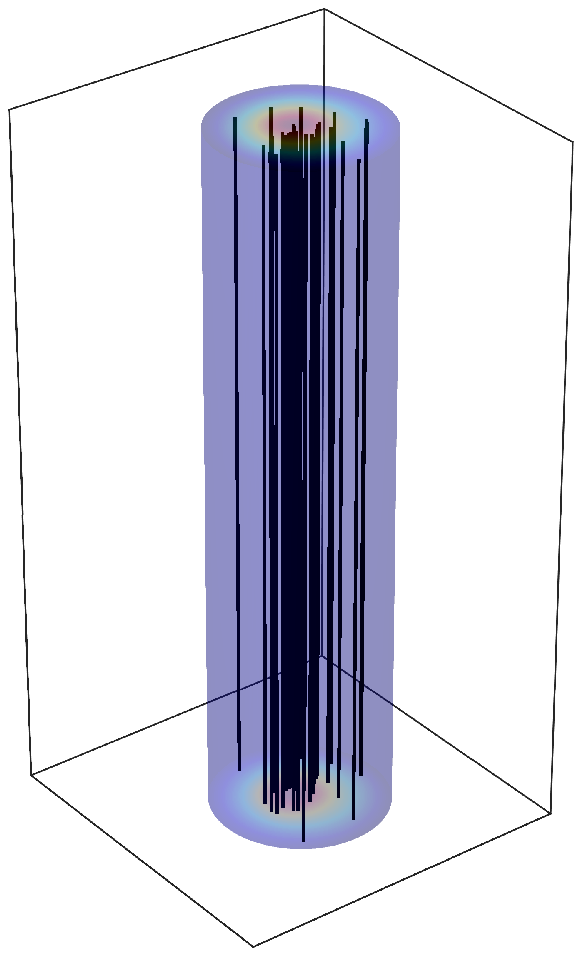}
\caption{Images of a $N=32$ vortex filament bundle of width $0.2D= 0.02~\rm cm$ surrounded by iso-surfaces of the coarse-grained vorticity field (left) and the coarse-grained negative pressure field (right). The positions of each vertical vortex filament are independently  sampled using a two-dimensional Gaussian distribution. The key observation is that both iso-surfaces are correlated and encapsulate the vortex bundle. \label{fig:bundle_tangle}}
\end{center}
\end{figure}

To study the result of coarse-graining of the flow fields, we first examine the simple case of a vertically orientated vortex bundle inside a periodic domain. We performed our analysis using a fixed sized bundle of radius $0.2D = 0.02~\rm cm$ consisting of Gaussian distributed vortex filaments numbering $N=8, 16, 32, 64$. Fig.~\ref{fig:bundle_tangle} displays the vortex bundle consisting of $N=32$ vortex filaments. The left image highlights the vorticity magnitude iso-surface after our coarse-graining procedure, while the right image shows a negative coarse-grained pressure iso-surface. Observe that the vortex bundle is encapsulated by the iso-surfaces indicating that the coarse-graining procedure is working. The coarse-graining procedure is as follows: by application of the vortex filament model~\eqref{eq:Schwarz-1}, we generate the superfluid velocity field $\mathbf{v}_{s}$ on a uniform three-dimensional spatial mesh of size $128\times 128\times 128$ which we then coarse-grain by applying a Gaussian low-pass filter $\hat{F}({\bf k})$ to the Fourier amplitudes of the superfluid velocity field, defined in Fourier space by 
\begin{align*}
\hat{F}(|{\bf k}|) = \exp\left(-\frac{|{\bf k}|^2}{2k_f^2}\right).
\end{align*}
The parameter $k_f = 2\pi/l_f$ represents the filtering wave length of the spatial filtering scale $l_f$ of the filter. Note, we explored several possible filtering algorithms, namely cubic spline, box, and Gaussian filters defined in physical space, with no substantial qualitative differences to that of the, vastly faster, spectral filter.  One can imagine that this filtering process represents the spatial resolution of an experimental probe limited to scales $l_f$ and above. Due to the irrotational nature of the flow, the superfluid vorticity field of the unfiltered velocity field is singular, hence this filtering process also acts as a natural regularization of the numerical data arising from the vortex filament model. We compute the coarse-grained pressure field using Eq.~\eqref{eq:pressure} applied to the filtered velocity field. A negative pressure iso-surface for the $N=32$ vortex bundle is shown in Fig.~\ref{fig:bundle_tangle} (right). For Fig.~\ref{fig:bundle_tangle}, we take the filtering scale $l_f$ to correspond to twice the mean inter-vortex spacing $\ell = \left(V/L_{\rm tot}\right)^{1/2}$ (where specifically for the bundle we have taken $V$ to be the bundle volume, and not the volume of the periodic box $V=D^3$). For the $N=32$ bundle, $\ell= 6.267\times 10^{-3}~\rm cm$. Justification for setting $l_f = 2\ell$ is explained later in this article, but relating the filter scale with the inter-vortex spacing scale is a natural choice as it will smooth the flow across neighbouring vortices leading to well-defined coarse-grained fields compatible with the HVBK equations. This agreement is highlighted by the direct correspondence between the vorticity and negative pressure iso-surfaces in Fig.~\ref{fig:bundle_tangle}.


\begin{figure}[htp!]
\begin{center}
\includegraphics[width=0.4\columnwidth]{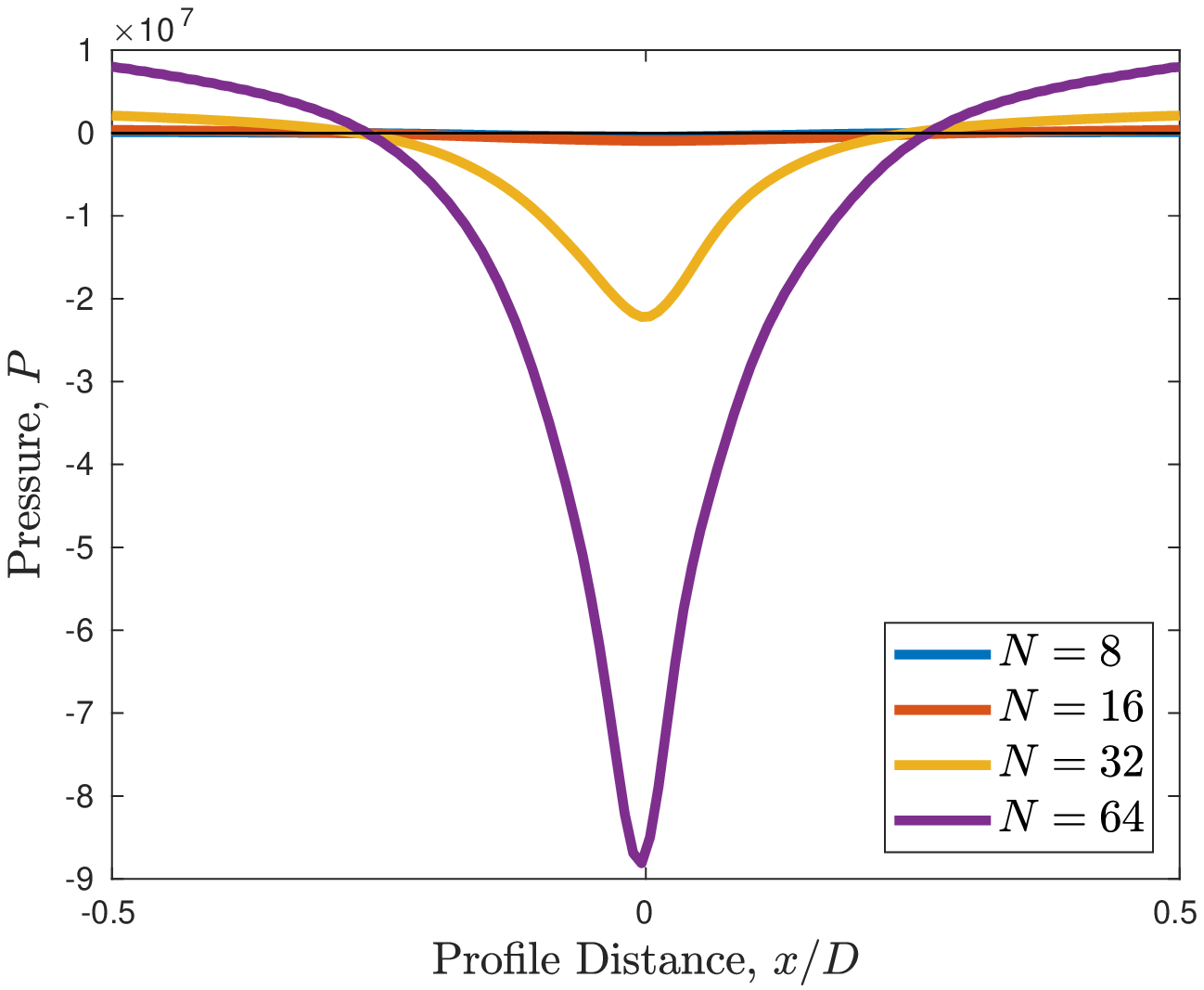}
\hspace{0.05\columnwidth}
\includegraphics[width=0.4\columnwidth]{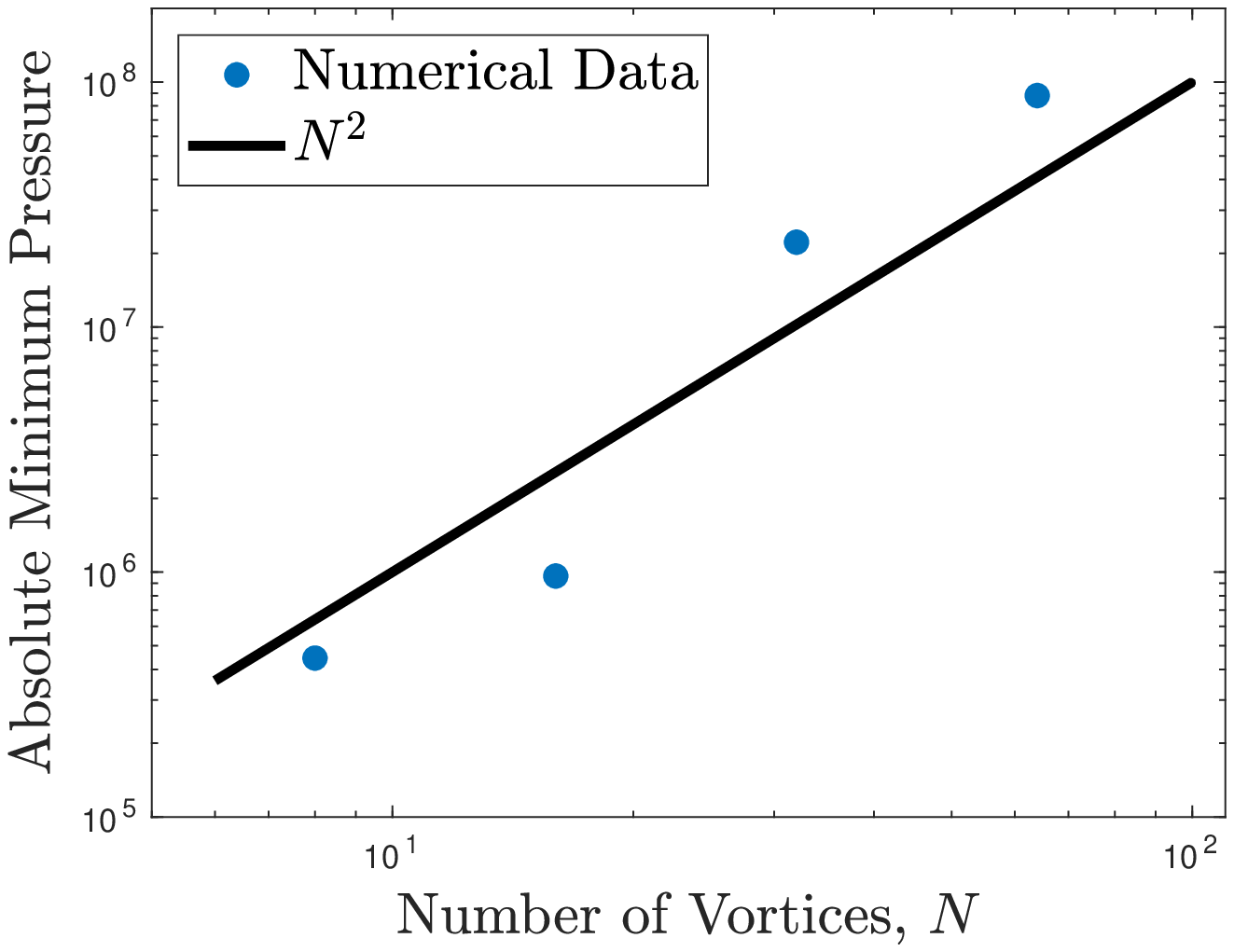}
\caption{(Left) Cross-sectional profile of the superfluid coarse-grained pressure field across the center of a vortex bundle composed of $N=8, 16 ,32, 64$ vertically aligned vortex filaments. We observe that by increasing the number of vortex filaments inside a fixed-sized bundle, one intensifies the negative pressure region located at the center of the vortex bundle. (Right) Observation of the course-grained peak negative pressure scaling~(\ref{eq:Pmin}) as the number of vortex filaments $N$ increases.\label{fig:bundle_pressure_profile}}
\end{center}
\end{figure}

In Fig.~\ref{fig:bundle_pressure_profile} (left) we plot the cross-sectional profile of the coarse-grained pressure field across the center of a vortex bundle composed of $N=8, 16, 32, 64$ vertical vortex filaments. We observe a significant decrease in the peak negative pressure at the center the bundle as the number of filaments increase. From plotting the peak negative pressure against the number of vortex filaments, we observe in Fig.~\ref{fig:bundle_pressure_profile} (right) that the peak negative pressure grows approximately as $\sim N^2$ in a fixed-sized bundle. The origin of the $P\sim N^2$ scaling can be determined by assuming that the bulk coarse-grained velocity field will have an azimuthal component 
\begin{align}\label{eq:bundle_vel}
v_\theta \sim \frac{N\Gamma}{2\pi r},
\end{align}
away from the bundle core. One may expect that due to the close interaction of multiple vortex filaments, the azimuthal component of velocity field inside the core of the vortex bundle could grow linearly with $r$ as in the Rankine vortex of classical hydrodynamics. However as pressure is a nonlocal quantity and requires full knowledge of the velocity field, we can expect that the majority of the pressure contribution arises from the  bulk velocity field away from the bundle core. Using this fact, we can directly substitute the bulk velocity field~\eqref{eq:bundle_vel} into Eq.~\eqref{eq:euler} (using cylindrical coordinates) with $\mathbf{F}_{\rm mf}=0$ and assuming stationarity of the flow. This permits direct calculation of the pressure $P$ giving 
\begin{align}\label{eq:Pmin}
P \sim P_0-\frac{\rho_s N^2 \Gamma^2}{8 \pi^2 r^2},
\end{align}
suggesting that the negative pressure would scale as $P\sim N^2$, as observed in Fig.~\ref{fig:bundle_pressure_profile}. The apparent smoothness of the pressure profile in Fig.~\ref{fig:bundle_pressure_profile} (left) is related to the inversion of the Laplace operator in the pressure computation leading to a naturally smoother field to that of the velocity, which is already filtered by the coarse-graining procedure. Reducing the filtering scale below our chosen value of $l_f=2\ell$ leads to the appearance of pressure fluctuations around the profile, possibly related to the unfiltered bundle core.


\section{Static Superfluid Tangles}

\begin{figure}[!htp]
\begin{center}
\includegraphics[width=0.6\textwidth]{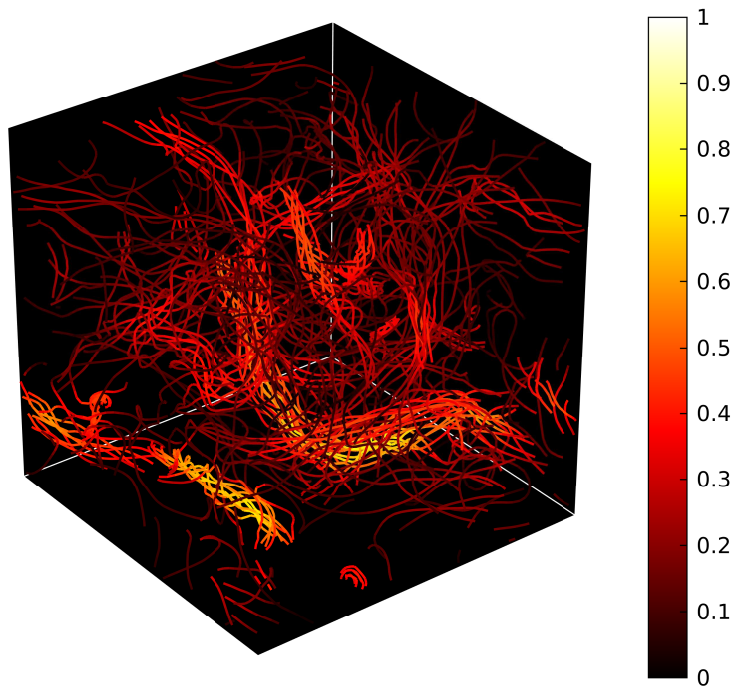}
\caption{Snapshot of the superfluid vortex tangle produced by vorticity locking to a classical Kolmogorov $k^{-5/3}$ turbulent normal fluid flow. The vortex filaments are colored by their local polarization $p$ of the local filaments~\cite{baggaley_vortex-density_2012}. The vortex line density $L = L_{\rm tot}/V$, where $L_{\rm tot}$ is the total line length and $V=0.1~{\rm cm}^3$ is the volume of the periodic box, is approximately $L\approx 2\times 10^{4}{\rm \ cm^{-2}}$. \label{fig:static_tangle}}
\end{center}
\end{figure}

The purpose of the single bundle analysis was to verify the coarse-graining procedure and to understand, from a simple setup, how pressure is related to the density of vortex filaments. We now proceed to examine more complex vortex configurations -- superfluid vortex tangles. We will consider two types of tangles consisting of (i) randomly positioned and orientated vortex rings mimicking a random or ultra-quantum tangle, and (ii) a quasi-classical superfluid tangle that is consistent with a Kolmogorov $k^{-5/3}$ (superfluid) energy spectrum.  
The latter is generated by running the vortex filament method coupled to a static turbulent normal fluid velocity field produced by the 3D Navier-Stokes equations with a non-zero mutual friction coupling, see~\cite{sherwin-robson_local_2015} for further details. As shown in Ref.~\cite{morris_vortex_2008}, evolving the superfluid velocity field while keeping the normal fluid component static results in {\it vorticity locking}, with quantized vortex filaments aligning to regions of high normal fluid vorticity leading to the superfluid field mimicking that of the normal fluid field. This produces a quasi-classical Kolmogorov $k^{-5/3}$ energy spectrum in the superfluid velocity field and the formation of coherent vortex bundles as can be observed in Fig.~\ref{fig:static_tangle}. The vortex tangle depicted in Fig.~\ref{fig:static_tangle} is colored by the local polarization $p$ of the vortex filaments highlighting regions of localized bundles that mimic the coherent vortex worms of classical turbulence. To compute the local polarization $p({\bf s}_i)$ at each vortex filament discretization point ${\bf s}_i$, we follow Ref.~\cite{baggaley_vortex-density_2012} and compute the magnitude of an effective local vorticity using a kernel with finite support, the $M_4$ kernel~\cite{monaghan_smoothed_1992}, which is effectively a cubic spline:
\begin{align*}
p({\bf s}_i) = \left|\kappa \sum_{j=1}^N\ {\bf s}'_j\ W(r_{ij},h)\ \Delta \xi_j \right|,
\end{align*}
where $r_{ij}=|{\bf s}_i-{\bf s}_j|$, $\Delta \xi_j = |{\bf s}_{j+1}-{\bf s}_j|$, $W(r,h)=g(r/h)/(\pi h^3)$, $h$ is the characteristic length scale which we set to be $h=2\ell$, and
\begin{align*}
g(q)=\begin{cases} 
1-\frac{3}{2}q^2 + \frac{3}{4}q^3 & 0 \leq q < 1,\\
\frac{1}{4}(2-q)^3 & 1\leq q < 2,\\
0 & 2\leq q. 
\end{cases}
\end{align*}
This approach is commonly employed in the smoothed particle hydrodynamics literature~\cite{monaghan_smoothed_1992}.

\begin{figure}[htp!]
\begin{center}
\includegraphics[width=0.4\columnwidth]{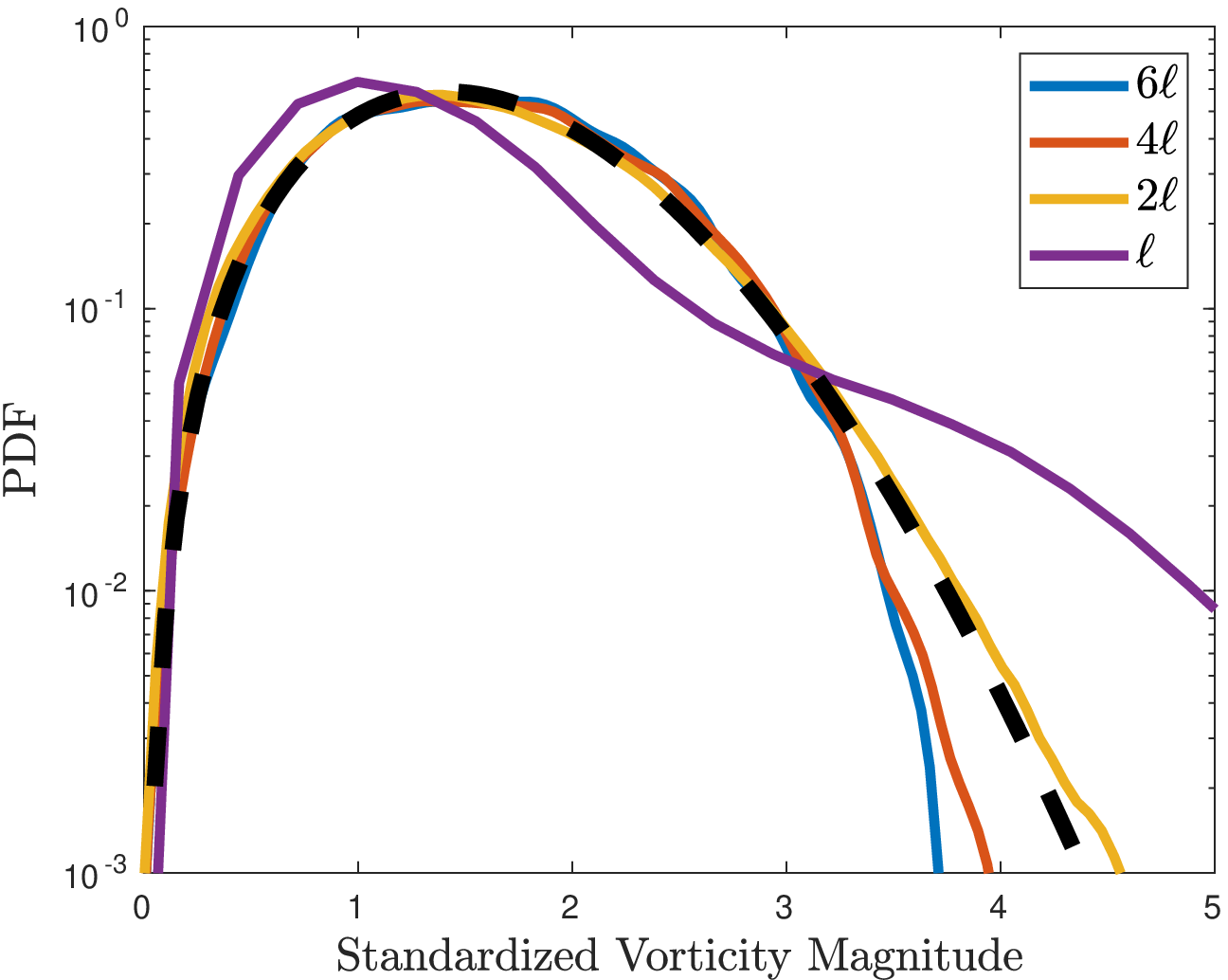}
\hspace{0.05\columnwidth}
\includegraphics[width=0.4\columnwidth]{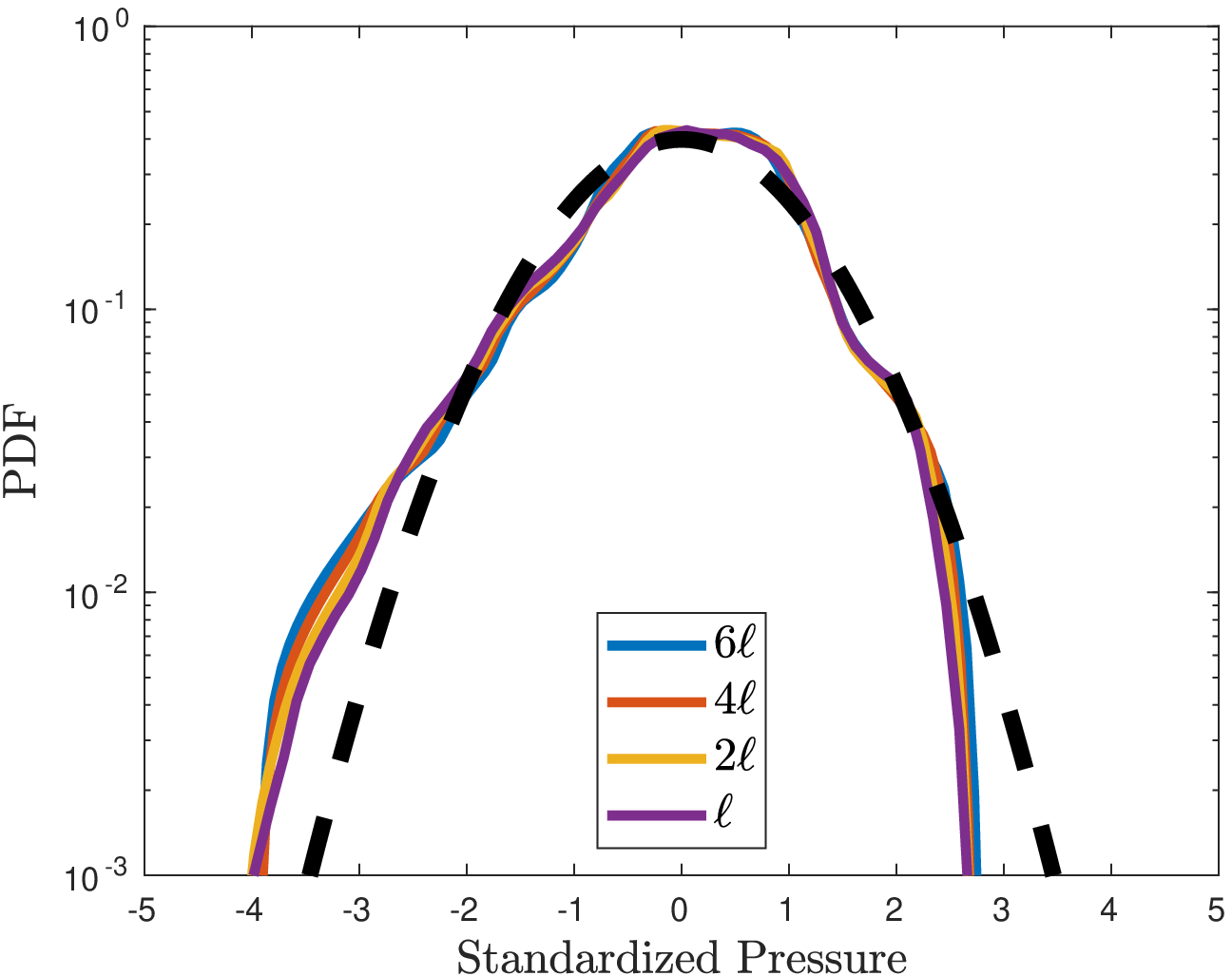}

\vspace{10pt}

\includegraphics[width=0.4\columnwidth]{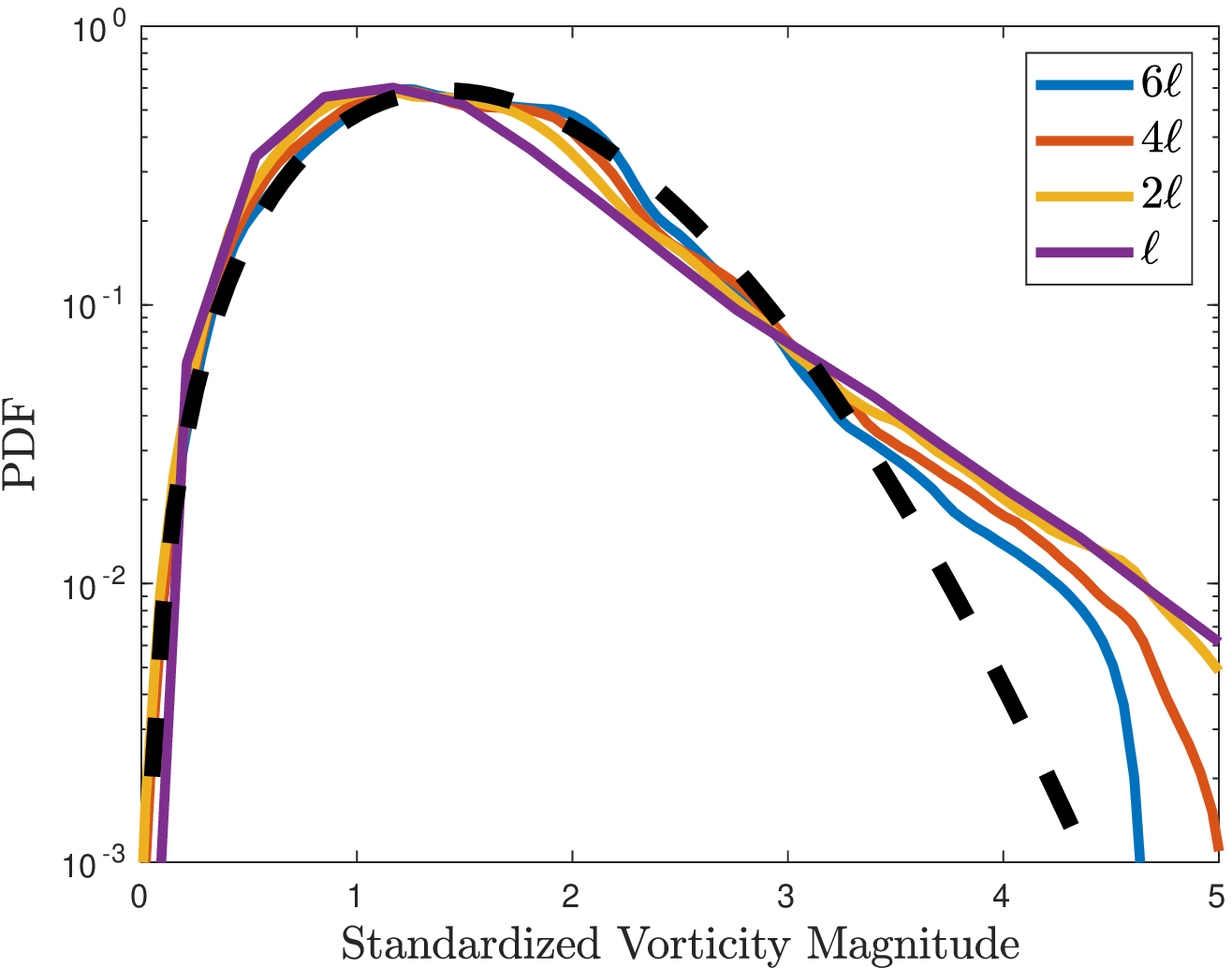}
\hspace{0.05\columnwidth}
\includegraphics[width=0.4\columnwidth]{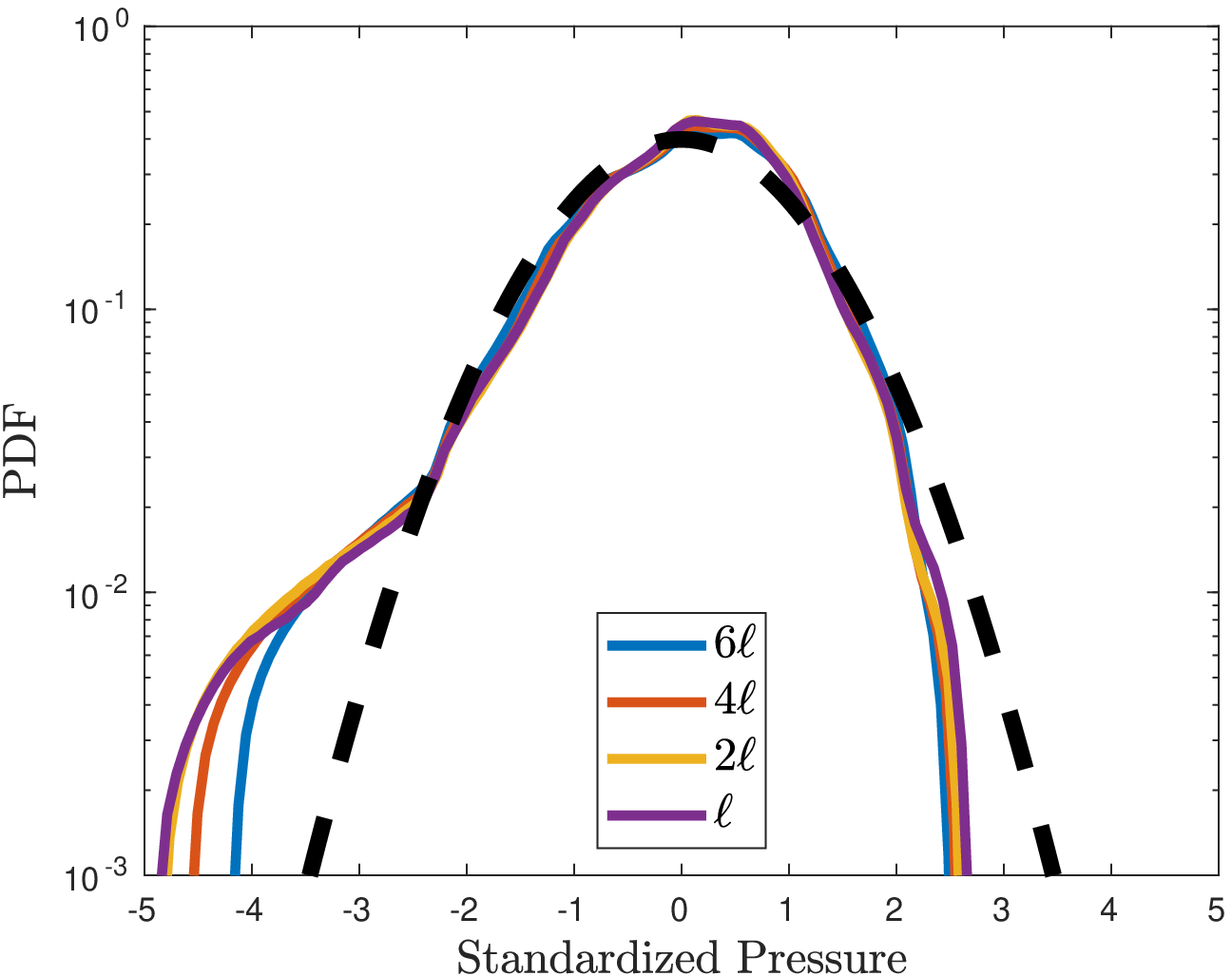}
\caption{PDFs of the standardized vorticity (left) and standardized pressure (right) for the random tangle (top) and quasi-classical (bottom) tangles with different coarse-graining parameters $l_f= \ell, 2\ell, 4\ell, 6\ell$, where $\ell = (V/L_{\rm tot})^{1/2}$ is the mean inter-vortex spacing. The black dashed curves indicate a pure Maxwellian distribution for the standardized vorticity (left) and Gaussian distribution for the standardized pressure (right). \label{fig:static_pdfs}}
\end{center}
\end{figure}

We probe the velocity and pressure fields of the static tangles using our coarse-graining procedure. In Fig.~\ref{fig:static_pdfs} we plot kernel density estimates of the probability density functions (PDFs) of the standardized vorticity magnitude $\omega/a_{\rm vort}$ (where $|\vect{\omega}|=\omega$) (left) and standardized pressure $P/\sigma_{\rm press}$ (the mean of the pressure is zero due to the periodicity of the domain) (right) of the random (top) and quasi-classical (bottom) tangles using different values of the filtering scale $l_f = \ell, 2\ell, 4\ell, 6\ell$. Parameter $a_{\rm vort}$ arises by assuming that for a three-dimensional random field with Gaussian statistics, the vorticity magnitude will have a Maxwell distribution with parameter $a$ defined as 
\begin{align}\label{eq:maxwell}
\mathbb{P}_{\rm Maxwellian}(x) = \sqrt{\frac{2}{\pi}}\frac{x^2\exp\left(-\frac{x^2}{2a^2}\right)}{a^3}, \quad a=\sqrt{\frac{\mathbb{E}[X^2]}{3}}.
\end{align}
For truly random fields the pressure field will have a Gaussian distribution, with PDF
\begin{align}\label{eq:gaussian}
\mathbb{P}_{\rm Gaussian}(x) =\frac{1}{\sqrt{2\pi\sigma^2}}\exp\left(-\frac{x^2}{2\sigma^2} \right),
\end{align}
where $\sigma$ is the standard deviation.

For the standardized vorticity magnitude, Fig.~\ref{fig:static_pdfs} (top and bottom, left), we observe clear indications of intermittency at high vorticity when $l_f=\ell$ which is subsequently removed by increasing the filtering or coarse-graining scale $l_f$. The vortex filament model generates singular velocity distributions due to the one-dimensional approximation of quantum vortex lines, meaning that high velocity (and therefore vorticity) values can appear in spatial mesh points located near vortex filaments. However, sufficient coarse-graining will circumvent this issue, and is clearly demonstrated when $l_f>\ell$ for the random tangle, Fig.~\ref{fig:static_pdfs} (top left). It seems reasonable to have a coarse-graining scale slightly larger than the average distance between adjacent vortex filaments in order to capture some large-scale effects, but not too large such that all medium-sized structures, like bundles, are filtered out. For the quasi-classical tangle, Fig.~\ref{fig:static_pdfs} (bottom left),  there are signs of strong high vorticity intermittency for all coarse-graining scales which relates to the probing of strong polarization in the form of coherent vortex bundles. Moreover, there are significant signs of negative pressure intermittency (enhancement of negative pressure over Gaussian statistics) at large negative pressures values in Fig.~\ref{fig:static_pdfs} (bottom right). This is significantly stronger than what is observed for the random tangle, Fig.~\ref{fig:static_pdfs} (top right), and is again related to the presence of coherent structures and the correlation between vorticity and pressure.

The authors of Ref.~\cite{rusaouen_detection_2017} conjectured that coherent vortex bundles could be identified by extreme negative pressure fluctuations exceeding $-3\sigma_{\rm press}$, which correspond to the strong intermittency tails of the negative pressure fields observed in Fig.~\ref{fig:static_pdfs}. Interestingly, when the coarse-graining scale $l_f$ increases we begin to reduce the intermittent effects and move towards a pure Maxwellian or Gaussian distribution of our fields. This is because the filter scale $l_f$ is approaching (or exceeding in the case of the random tangle) the length scale associated to the largest vortex bundle size. For the random tangle, there are intrinsically no bundles, so the PDFs converge relatively quickly as $l_f>\ell$, while for the quasi-classical tangle, we observe that the typical vortex bundle size is approximately $\sim 6\ell$ or larger, see Fig.~\ref{fig:static_tangle}, so even for $l_f=6\ell$ not all the coherent structures are filtered. This is an important observation when analyzing experimental data, and suggests that the data presented in Ref.~\cite{rusaouen_detection_2017} indicates that the embedded coherent structures in their experimental flow are larger than the probe scale. As the dynamics of the flow dictate the vortex line density $L$, and subsequently the mean inter-vortex spacing $\ell = (V/L_{\rm tot})^{1/2} = L^{-1/2}$,  it is important for us to choose a dynamic filtering scale $l_f$ to ensure that coherent structures are not averaged out during turbulence decay. Therefore, to be consistent, we select the coarse-graining scale to be twice the mean inter-vortex spacing $l_f=2\ell$ in our analysis.

\begin{figure*}[!htp]
\begin{center}
\includegraphics[width=0.4\textwidth]{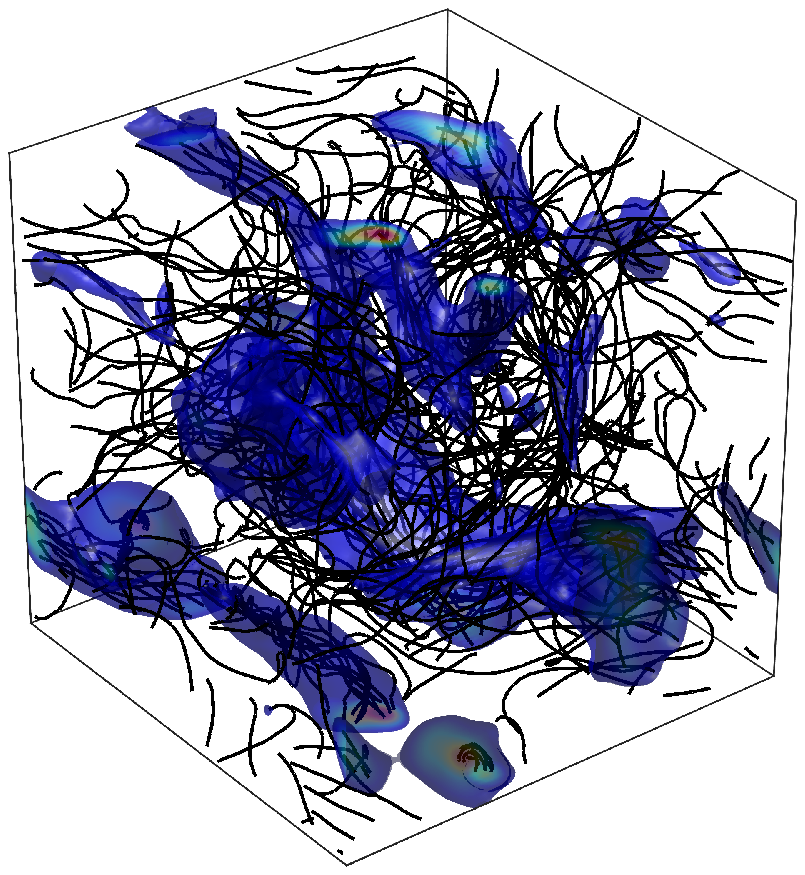}
\hspace{0.05\columnwidth}
\includegraphics[width=0.4\textwidth]{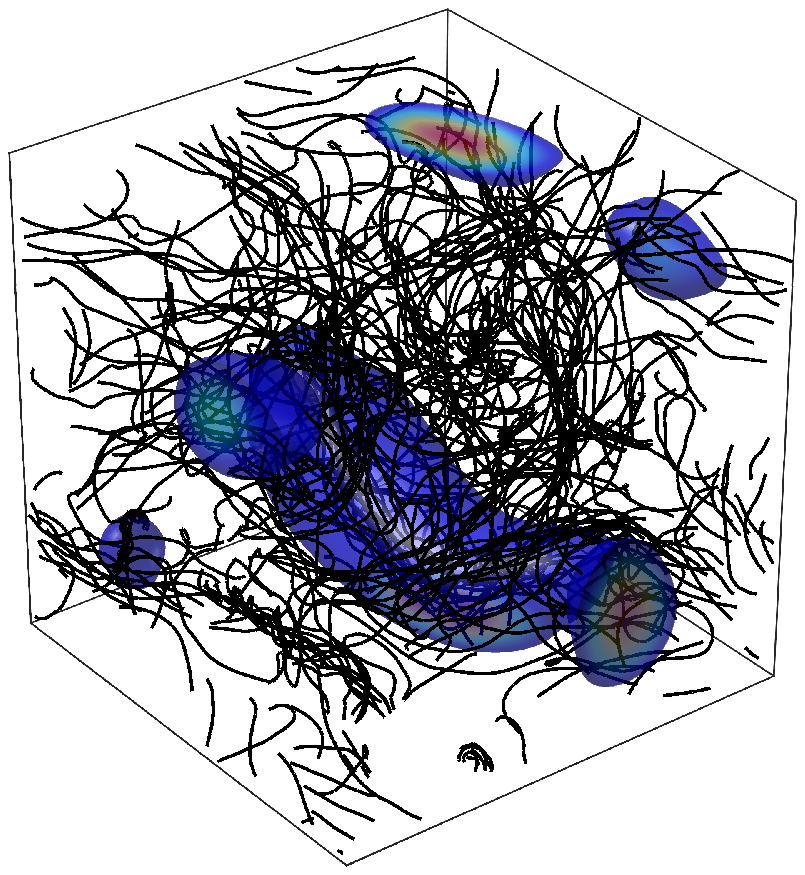}
\caption{Iso-surfaces of the coarse-grained standardized vorticity magnitude (left) and negative pressure (right) fields of the static quasi-classical tangle depicted in Fig.~\ref{fig:static_tangle} using $l_f=2\ell$. The iso-surfaces are taken at $\omega/a_{\rm vort} > 2.5$ and $P/\sigma_{\rm press} < -1.5$ respectively. \label{fig:static_iso} }
\end{center}
\end{figure*}

To further verify that setting $l_f=2\ell$ is reasonable, we display iso-surfaces of coarse-grained vorticity magnitude and negative pressure fields of the quasi-classical tangle in Fig.~\ref{fig:static_iso}. We observe clear correlation between the intense coarse-grained vorticity regions encompassing the regions of strong locally polarized vortex bundles, see Fig.~\ref{fig:static_tangle} and Fig.~\ref{fig:static_iso} (left), which are also correlated to regions of high negative pressure Fig.~\ref{fig:static_iso} (right).

\begin{figure}[htp!]
\begin{center}
\includegraphics[width=0.45\textwidth]{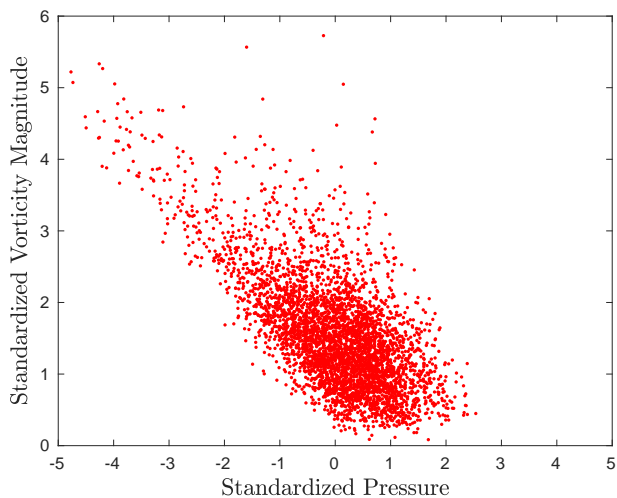}
\caption{Scatter-plot of the standardized coarse-grained vorticity magnitude and pressure fields of the static quasi-classical tangle depicted in Fig.~\ref{fig:static_tangle}. The standardizing parameters are $a_{\rm vort} = 5.89\times 10^4$ and $\sigma_{\rm press}= 4.83\times 10^5$. \label{fig:static_scatter} }
\end{center}
\end{figure}

To quantify the relationship between the vorticity and pressure, numerical simulations allow for simultaneous measurements of both fields across the computational  domain. In Fig.~\ref{fig:static_scatter} we present a scatter-plot of the standardized vorticity magnitude versus the standardized pressure of the quasi-classical tangle. A clear trend between regions of strong absolute vorticity and extreme negative pressure is visible. The most extreme of which are located in regions of $P/\sigma_{press} \lesssim -3$  and $\omega/a_{\rm vort} \gtrsim 3.5$. This provides further evidence that extreme negative pressure fluctuations are indications of large vorticity regions produced at the center of coherent vortex bundles.

\section{Decaying Turbulence}

Whilst the experiments of Rusaouen {\it et al.} \cite{rusaouen_detection_2017} focused on statistical properties of turbulence in a non-equilibrium steady-state regime, there is also merit to studying properties of the decay of turbulence. Indeed, in experimental studies of superfluid turbulence this can often elucidate important properties which cannot be directly measured in the steady-state~\cite{walmsley_quantum_2008,babuin_quantum_2012}. Thus, having investigated the properties of a static tangle, we now investigate the temporal dynamics of three different structures of superfluid turbulence tangles. These are: (i) a random tangle with no structure, (ii) an imposed large-scale Taylor-Green flow vortex tangle, and finally (iii) a quasi-classical Kolmogorov tangle. Each simulation is performed in the low-temperature limit with $\alpha=0.01$ and $\alpha'=0$ which corresponds to a temperature $T\lesssim 1.1~{\rm K}$ were the two-fluid description implies $\rho_s/\rho \simeq 0.99$.

We utilize the random and quasi-classical tangles from before as initial conditions. For the Taylor-Green simulation, we follow Refs.~\cite{nore_decaying_1997,araki_energy_2002} by taking an initial condition that consists of a series of vortex filaments that follows the classical Taylor-Green velocity field. This produces a predominately large scale flow of the order of the box size. Each initial condition has approximately the same vortex line density of $L \approx 2\times 10^4~ {\rm cm}^{-2}$ so comparisons can be drawn. We evolve all three tangles to a fully developed turbulence state to ensure any artificial transient features arising from the initial conditions are absent. 

\begin{figure}[htp!]
\begin{center}
\includegraphics[width=0.45\textwidth]{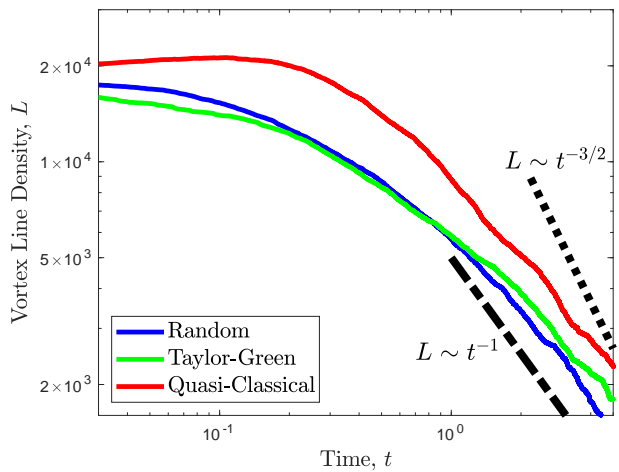}
\caption{Evolution of the vortex line density $L$ of the random (blue), Taylor-Green (green), and quasi-classical (red) tangles. The long-time decay of all the tangles scale well with the $L\sim t^{-1}$ scaling expected for random tangles. Surprisingly, the decay observed in the structured tangles (Taylor-Green and quasi-classical) are shallower than the expected $L\sim t^{-3/2}$ scaling. This is probably due to a lack of scale separation between the box-size $D$ and inter-vortex spacing $\ell$ at late times of our simulations. \label{fig:vld}}
\end{center}
\end{figure}

As the system evolves the vortex line density decays as our systems are unforced. The functional relationship between $L$ and $t$ in decaying quantum turbulence is an important indicator of the underlying nature of the flow. In the quasi-classical regime the long-time decay is expected to follow a power-law scaling, $L \sim t^{-3/2}$, whereas the unstructured random tangle should decay at long-times as $L \sim t^{-1}$~\cite{walmsley_dynamics_2014}. The difference between these scalings arises due to the hierarchy of hydrodynamic scales present in the quasi-classical regime, which are absent in the essentially single-scaled random tangle.
Fig.~\ref{fig:vld} shows the decay of the vortex line density $L$ versus time for all three tangles. In all three, the decay is constant with $L \sim t^{-1}$, suggesting that as the system evolves it becomes dominated by a single scale, the inter-vortex spacing $\ell$. In the case of the Taylor-Green and quasi-classical initial conditions this is perhaps surprising, but is probably a symptom of limited scale separation between the box-size $D$ and $\ell$ in our numerical simulations. We also note that the moderate initial increase of the vortex line density in the quasi-classical tangle is a well-known phenomenon and is associated to the initial generation of Kelvin-waves as the tangle initially relaxes. 

\begin{figure}[htp!]
\begin{center}
\includegraphics[width=\textwidth]{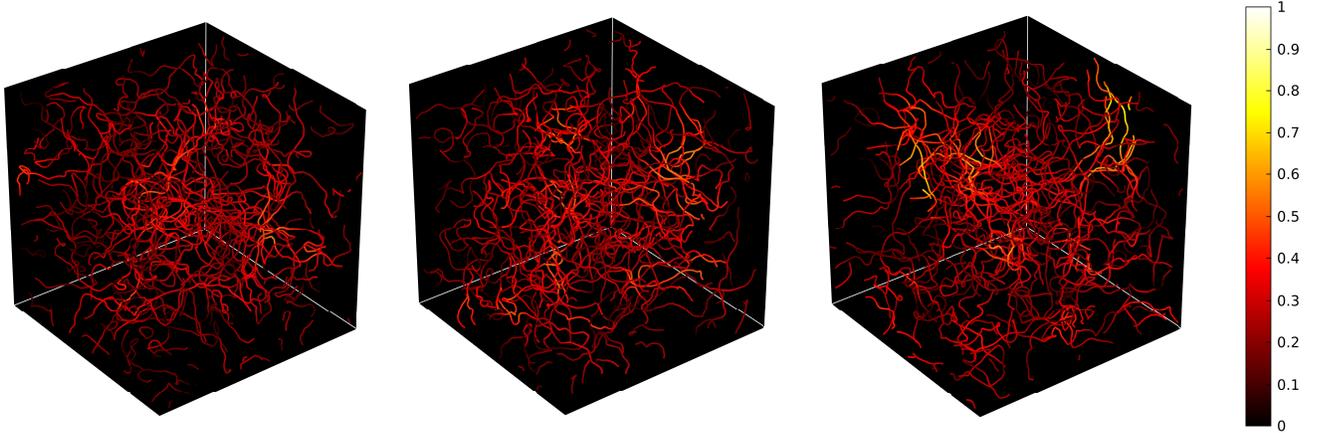}
\caption{Snapshots of the random (left), Taylor-Green (center), quasi-classical (right) vortex tangles at times where the vortex line density  $L=1\times 10^{4}~{\rm cm^{-2}}$. The vortex filaments are colored by the local polarization with the color scheme normalization to the maximum of the polarization across all three tangles. \label{fig:decay_tangle} }
\end{center}
\end{figure}

We examine the three tangles at the point in which each of their vortex line densities reach $L = 1\times 10^4~{\rm cm}^{-2}$. Table~\ref{table:data} displays the main statistics of the three tangles at this moment in their evolution. Snapshots of the three tangles at this time are presented in Fig.~\ref{fig:decay_tangle}, with the vortex filaments colored according to their local polarization $p$. The color range is normalized to the maximum polarization across the three tangle which occurs in the quasi-classical tangle (far right). Observe how in the quasi-classical tangle the presence of coherence structures have greatly reduced compared to those observed in the initial condition of Fig.~\ref{fig:static_tangle}, but are nonetheless still observed. Interestingly, there is some weak polarization in both the Taylor-Green and random tangles, with the latter having the weakest polarization as expected.

\begin{table}
\begin{center}
\caption{Key statistics of the decaying simulations of the three types of superfluid vortex tangle. $v_{\rm rms}$ is the random mean square of the velocity field, $\bar{\omega}_{\rm mean}$ the mean of the absolute vorticity field, $\sigma_{\rm vel}$, $\sigma_{\rm press}$, $\sigma_{\rm vort}$ are the standard deviations of the absolute velocity, pressure, and absolute vorticity fields respectively, and $a_{\rm vel}$, $a_{\rm vort}$ are the scaled square root of the second moment of the absolute velocity and vorticity fields as defined in Eq.~\eqref{eq:maxwell}.\label{table:data}}
\begin{tabular}{c|ccccccc}
Type & $v_{\rm rms}$ & $\bar{\omega}_{\rm mean} $ &  $\sigma_{\rm vel}$  &  $\sigma_{\rm press}$ & $\sigma_{\rm vort}$ & $a_{\rm vel}$ & $a_{\rm vort}$\\
\hline
Random & $265.43$ &  $2.14\times 10^{4}$    & $83.42$      &   $1.74\times 10^{4}$ & $8.89\times 10^{3}$ &  $1.53\times 10^2$ &$1.34\times 10^{4}$\\
Taylor-Green & $424.76$ & $2.52\times 10^{4}$  & $188.68$   &   $4.03\times 10^{4}$ &   $1.24\times 10^{4}$ &  $2.45\times 10^2$ & $1.62\times 10^{4}$\\
Quasi-Classical & $425.33$ &  $2.73\times 10^{4}$     & $151.22$        &   $4.95\times 10^{4}$ &   $1.31\times 10^{4}$ & $2.46\times 10^2$ & $1.75\times 10^{4}$
\end{tabular}
\end{center}
\end{table}

\begin{figure*}[htp!]
\begin{center}
\includegraphics[width=0.3\textwidth]{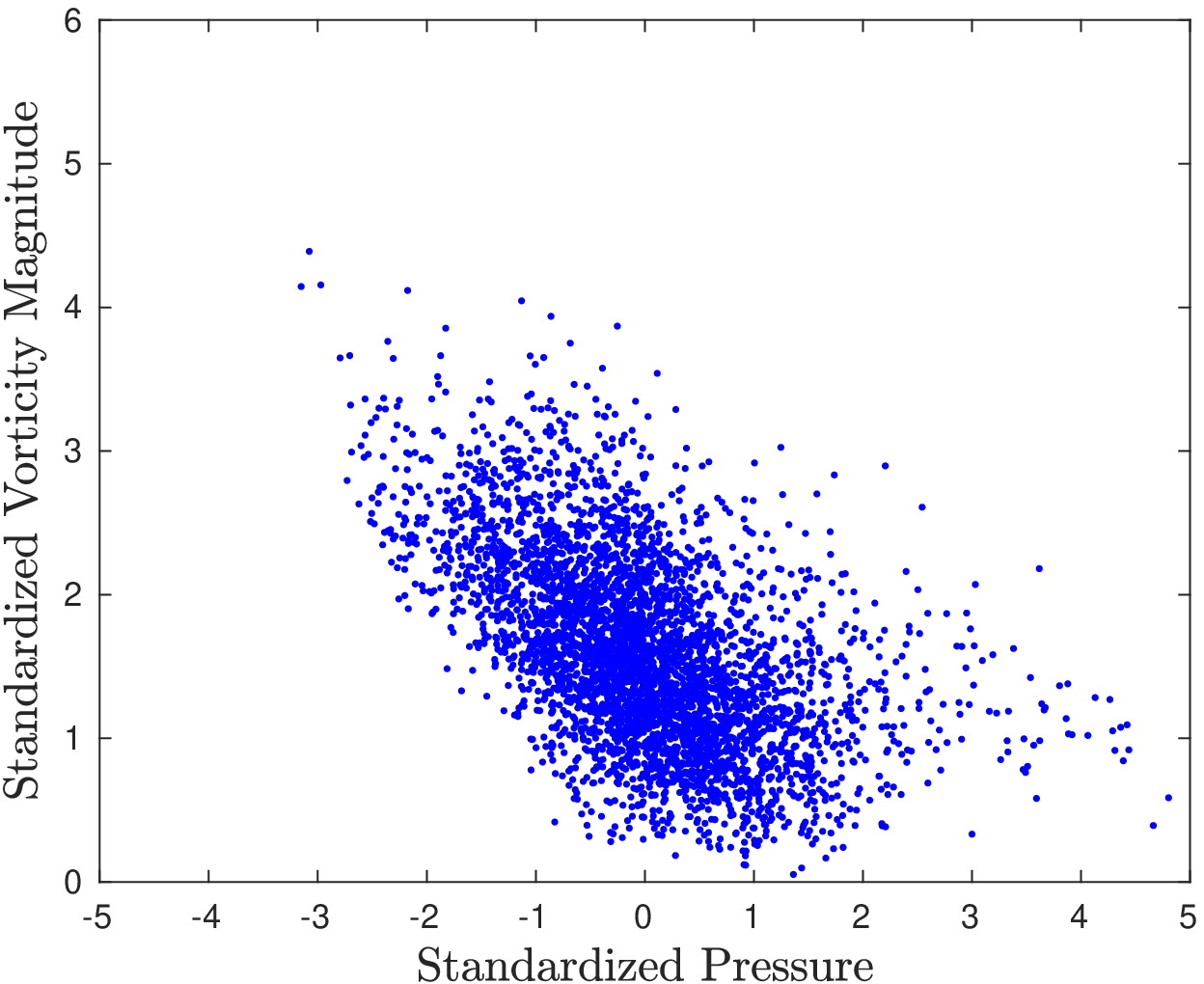}
\hspace{0.02\columnwidth}
\includegraphics[width=0.3\textwidth]{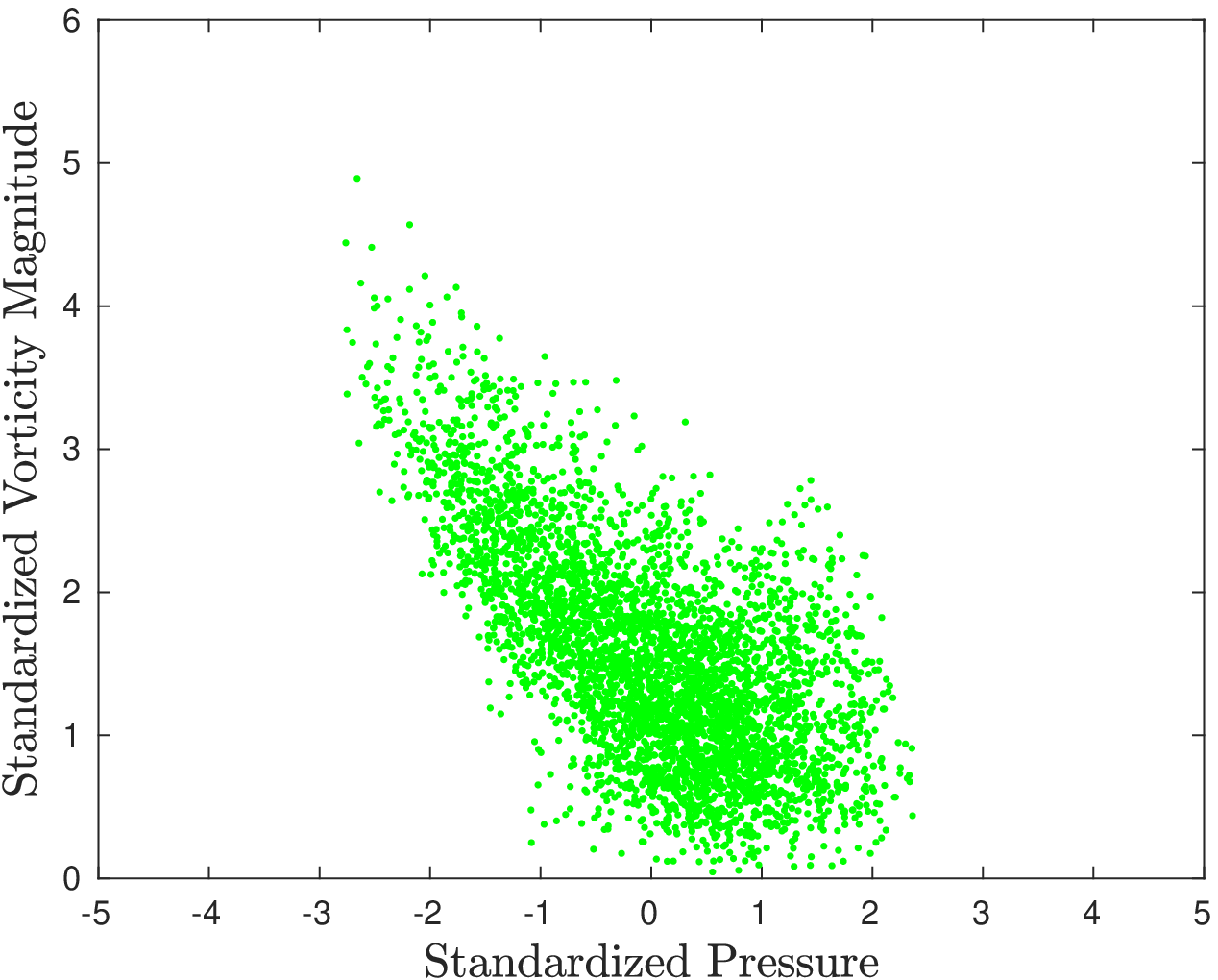}
\hspace{0.02\columnwidth}
\includegraphics[width=0.3\textwidth]{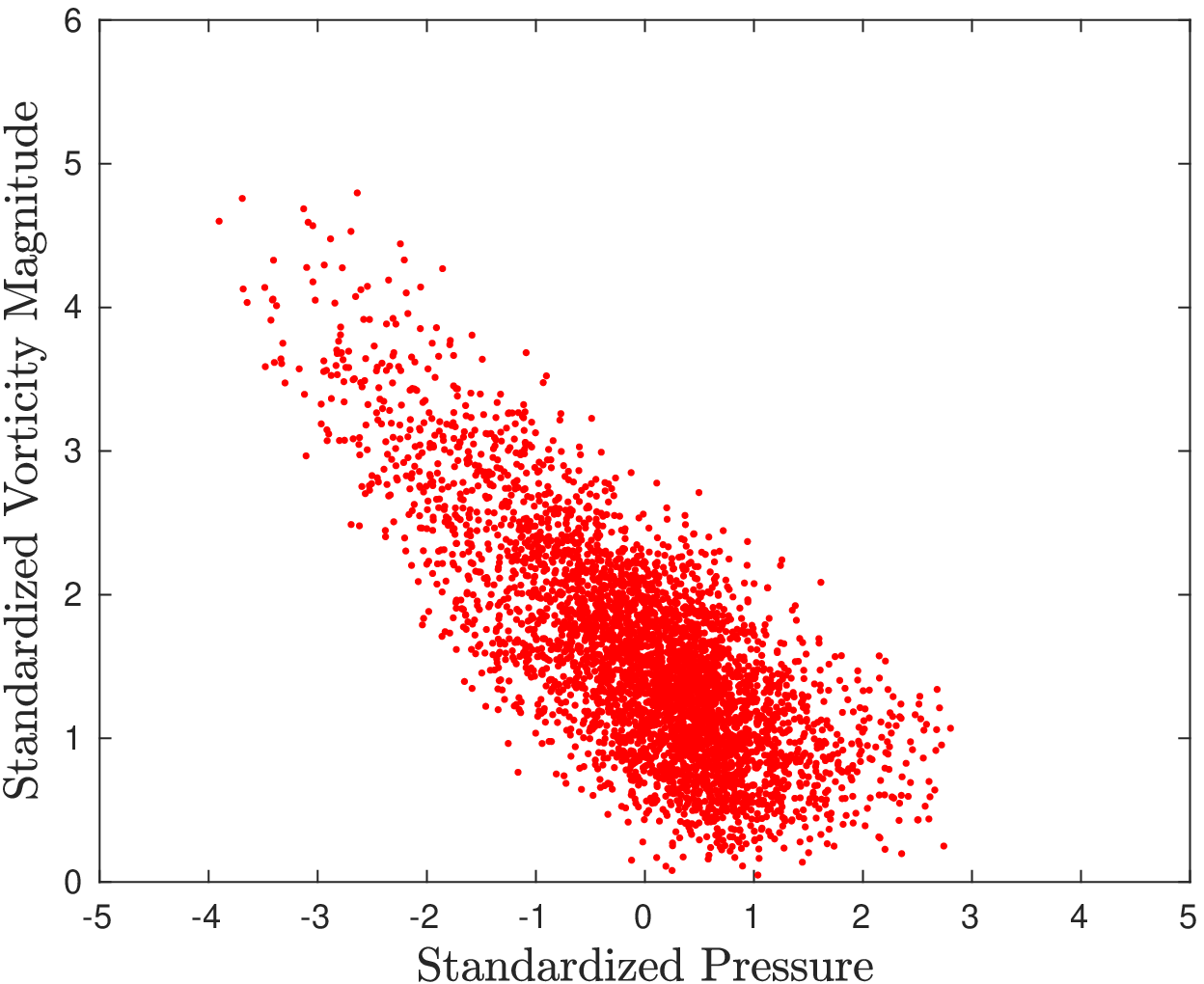}
\caption{Scatter-plots of the coarse-grained pressure and vorticity magnitude fields of the  random (left), Taylor-Green (center), quasi-classical (right) vortex tangle  at $L=1\times10^{4}~{\rm cm^{-2}}$.\label{fig:decay_scatter}}
\end{center}
\end{figure*}

Fig.~\ref{fig:decay_scatter} shows scatter-plots of the standardized coarse-grained vorticity magnitude versus standardized pressure for the three evolved vortex tangles: random (left), Taylor-Green (center), quasi-classical (right). In all three instances, we observe clear trends of large negative pressure regions being associated with large absolute vorticity that ends approximately when the negative pressure reaches values close to $-3\sigma_{\rm press}$. Observe that the negative pressure extremes only really exceeds the value $-3\sigma_{\rm press}$ for the quasi-classical tangle that still appear to contain many coherent structures. It is interesting to note that there is little variation in the distribution of the data across the three tangle structures, although the random tangle does appear to produce quite a significant proportion of high pressure regions. However, it is important to recall that we are displaying standardized quantities in Fig.~\ref{fig:decay_scatter}, and that the normalization variables $a_{\rm vort}$ and $\sigma_{\rm press}$ are clearly larger for the quasi-classical and Taylor-Green tangles.

\begin{figure}[htp!]
\begin{center}
\includegraphics[width=0.3\columnwidth]{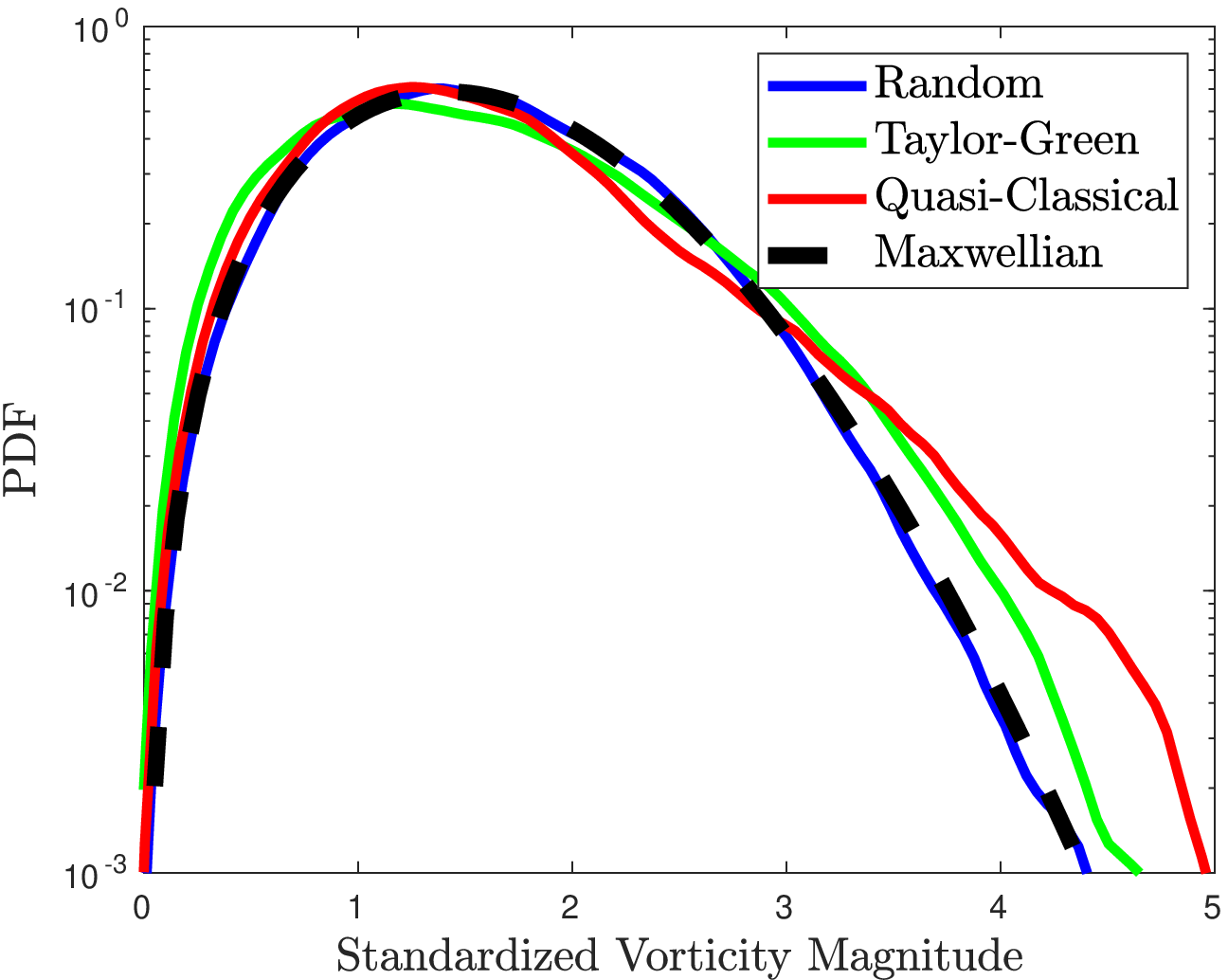}
\hspace{0.02\columnwidth}
\includegraphics[width=0.3\columnwidth]{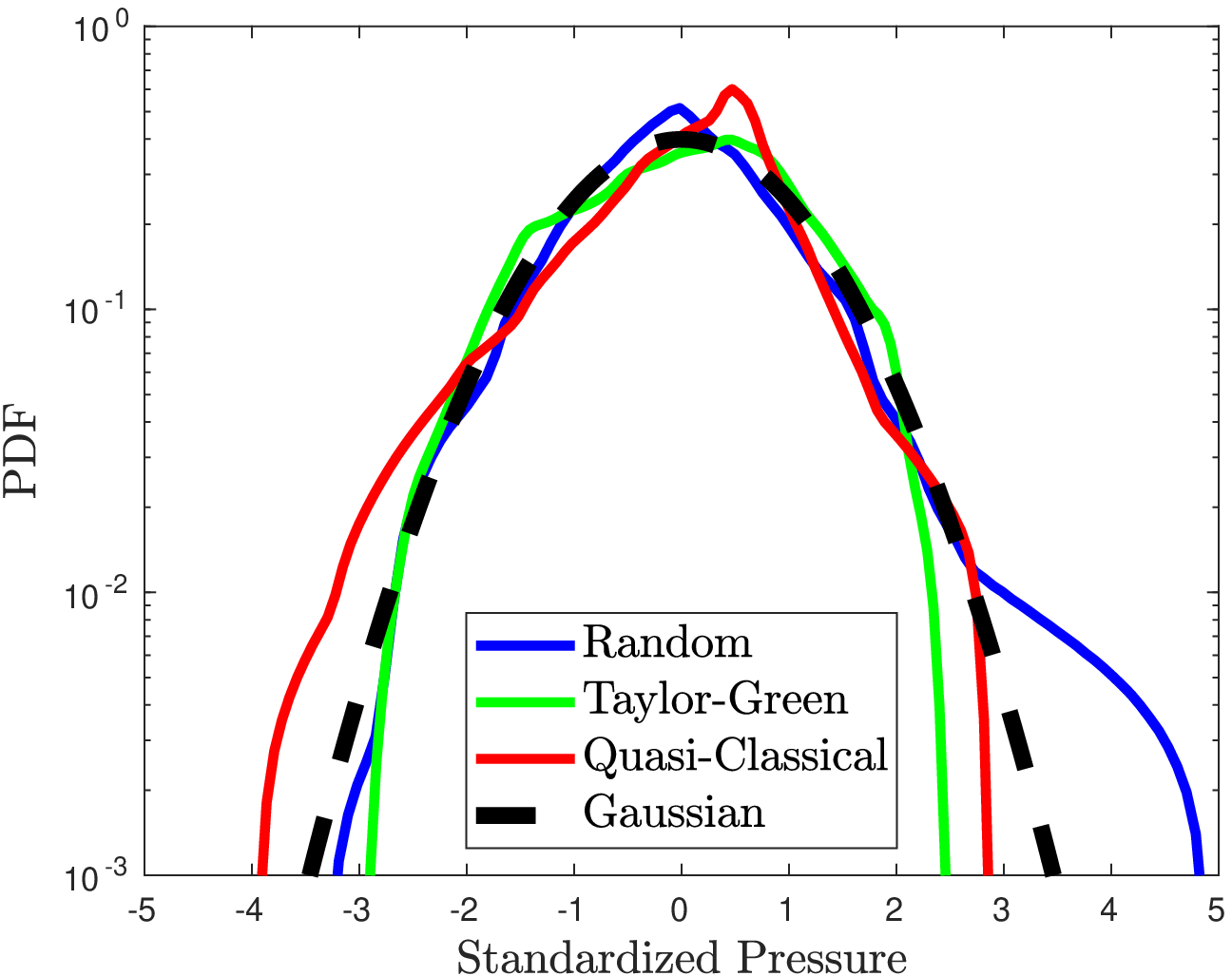}
\hspace{0.02\columnwidth}
\includegraphics[width=0.3\columnwidth]{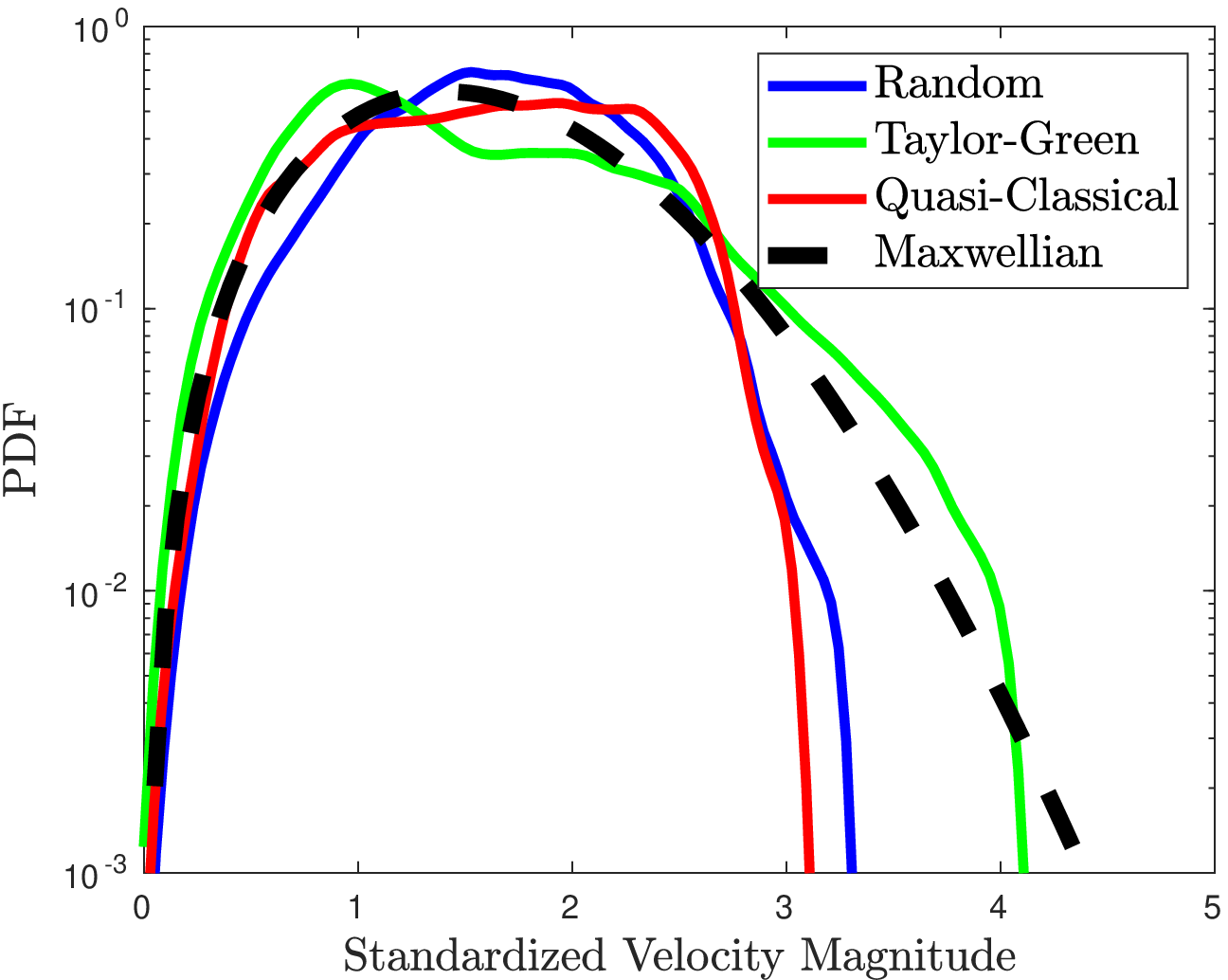}
\caption{Kernel density estimates of the PDFs of the standardized vorticity magnitude (left), standardized pressure (center), and standardized velocity magnitude (right) fields of the random (blue), Taylor-Green (green), and quasi-classical (red) tangles. The vorticity and velocity magnitudes PDFs are compared to the standardized Maxwellian distribution, while the pressure PDF is compared to the standardized Gaussian distribution. \label{fig:decay_pdf}}
\end{center}
\end{figure}

\begin{figure}[htp!]
\begin{center}
\includegraphics[width=0.45\columnwidth]{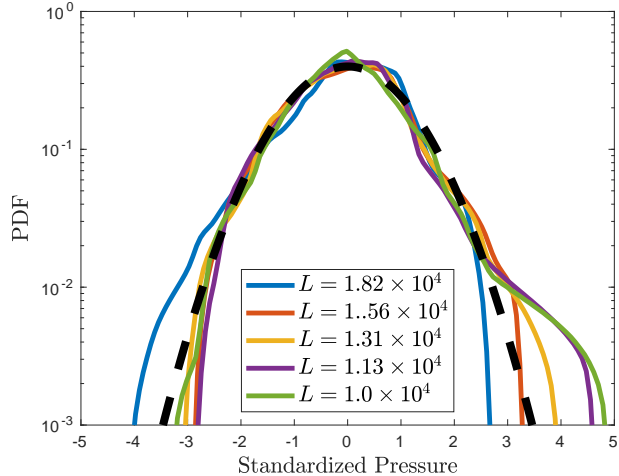}
\caption{Standardized pressure PDFs of the random tangle during decay evolution indicated by the value of the vortex line density $L$. Observe the enhancement of high pressure region of the PDF during the decay. \label{fig:random_pressure_pdf}}
\end{center}
\end{figure}

In an attempt to try and distinguish the types of tangles through the flow statistics, we compute the the PDFs of the standardized vorticity magnitude, pressure and velocity magnitude that are displayed in Fig.~\ref{fig:decay_pdf}. A key observation in Fig.~\ref{fig:decay_pdf} is that the PDFs show very little intermittency which is probably a consequence of the lack of many strong coherent structures as can be observed in the physical tangles presented in Fig.~\ref{fig:decay_tangle}. With that being said, we observe some slight enhancement of negative pressure in the quasi-classical tangle, while there is a significant deviation from Gaussianity for positive pressure in the random tangle. To understand this particular feature of the random tangle we refer the reader back to the approximate form of the mutual friction $\mathbf{F}_{\rm mf}$, Eq.~\eqref{eq:F_mf}, which implies a faster decay of the flow in regions of strong vorticity. For tangles with strong polarization, we would expect that the mutual friction would lead to enhanced dampening of strong vorticity and negative pressure regions, reducing intermittency. However, the pressure field of the random tangle is initially close to Gaussian, so the mutual friction may lead to relative enhancement of the high pressure regions due to the preferentially dampening of regions of low pressure, and ultimately leading to a skewing of the pressure field towards positive regions during its decay. This is confirmed in Fig.~\ref{fig:random_pressure_pdf} where we measure the standardized pressure PDF of the random tangle throughout the decay and observe the manifestation of this effect in a growing `bump' of high pressure. (Note that during the decay, the PDFs are continuously re-standardized, and the high pressure is not technically appearing but simply dominating the PDF.) It seems of interest that whilst the original motivation for monitoring the pressure in superfluid turbulence was to find a signal of the quasi-classical regime through the detection of coherent structures, it can also be of use in determining the structure of the ultra-quantum regime.

During the decay of the structured tangles we observe  a relaxation towards Gaussian statistics (not shown). This does not contradict the picture of dissipation acting primarily in regions of strong vorticity and the overall effect of increasing the skewness of the pressure statistics, as the initial conditions of these tangle displayed strong intermittency from the start in the form of heavy PDF tails in the high vorticity magnitude and negative pressure areas due to the presence of large-scale coherent structures.

\section{Conclusions}

In this work we have investigated coherent vorticity structures in superfluid turbulence, connecting the dynamics of individual quantized vortices to the macroscopic HVBK equations. In line with the conclusions of Rusaouen {\it et al.}~\cite{rusaouen_detection_2017}, we show that strong pressure drops in superfluid turbulence can be directly associated with coherent bundles of macroscopic vorticity, which have long been discussed as crucial for the observed quasi-classical behavior of quantum turbulence in many experimental and numerical studies. 

Across a series of numerical simulations we demonstrate strong correlations between negative pressure and vorticity, with the presence of vortex bundles leading to intermittency of both fields perturbing the underlying Maxwellian (vorticity) and Gaussian (pressure) distributions. Furthermore, we have shown a new high pressure bump emerging in decaying simulations of a purely random tangle that models the ultra-quantum regime, which is associated to the faster decay of high vorticity regions in the turbulence and could be used in future experimental studies of this regime. 

The conclusions of this article remain valid for the coldest experiments of Ref.~\cite{rusaouen_detection_2017}, performed at a temperature of $T=1.58K$, which gives the superfluid ratio at $83\%$. This is because the pressure $P$ should still be dominated by the superfluid component and the approximation going from Eq.~\eqref{eq:pressure_full} to Eq.~\eqref{eq:pressure} will remain reasonable. However, the question regarding at what temperatures should the normal fluid contribution to the pressure be taken into account is very intriguing and is worthy of a future study.

\bibliographystyle{ieeetr}
\bibliography{bibliography}

\begin{thebibliography}{10}

\bibitem{roche_triggering_2010}
P.-E. Roche, F.~Gauthier, R.~Kaiser, and J.~Salort, ``On the triggering of the
  {Ultimate} {Regime} of convection,'' {\em New Journal of Physics}, vol.~12,
  p.~085014, Aug. 2010.

\bibitem{rousset_superfluid_2014}
B.~Rousset, P.~Bonnay, P.~Diribarne, A.~Girard, J.~M. Poncet, E.~Herbert,
  J.~Salort, C.~Baudet, B.~Castaing, L.~Chevillard, F.~Daviaud, B.~Dubrulle,
  Y.~Gagne, M.~Gibert, B.~H{\'e}bral, T.~Lehner, P.-E. Roche, B.~Saint-Michel,
  and M.~B. Mardion, ``Superfluid high {REynolds} von {K{\'a}rm{\'a}n}
  experiment,'' {\em Review of Scientific Instruments}, vol.~85, p.~103908,
  Oct. 2014.

\bibitem{tisza_theory_1947}
L.~Tisza, ``The {Theory} of {Liquid} {Helium},'' {\em Physical Review},
  vol.~72, pp.~838--854, Nov. 1947.

\bibitem{landau_theory_1941}
L.~Landau, ``Theory of the {Superfluidity} of {Helium} {II},'' {\em Physical
  Review}, vol.~60, pp.~356--358, Aug. 1941.

\bibitem{landau_theory_1949}
L.~Landau, ``On the {Theory} of {Superfluidity},'' {\em Physical Review},
  vol.~75, pp.~884--885, Mar. 1949.

\bibitem{goto_physical_2008}
S.~Goto, ``A physical mechanism of the energy cascade in homogeneous isotropic
  turbulence,'' {\em Journal of Fluid Mechanics}, vol.~605, pp.~355--366, June
  2008.

\bibitem{salort_energy_2012}
J.~Salort, B.~Chabaud, E.~L{\'e}v{\^e}que, and P.-E. Roche, ``Energy cascade
  and the four-fifths law in superfluid turbulence,'' {\em EPL (Europhysics
  Letters)}, vol.~97, p.~34006, Feb. 2012.

\bibitem{laurie_interaction_2010}
J.~Laurie, V.~S. L{\textquoteright}vov, S.~Nazarenko, and O.~Rudenko,
  ``Interaction of {Kelvin} waves and nonlocality of energy transfer in
  superfluids,'' {\em Physical Review B}, vol.~81, p.~104526, Mar. 2010.

\bibitem{baggaley_kelvin-wave_2014}
A.~W. Baggaley and J.~Laurie, ``Kelvin-wave cascade in the vortex filament
  model,'' {\em Physical Review B}, vol.~89, p.~014504, Jan. 2014.

\bibitem{galantucci_crossover_2019}
L.~Galantucci, A.~W. Baggaley, N.~G. Parker, and C.~F. Barenghi, ``Crossover
  from interaction to driven regimes in quantum vortex reconnections,'' {\em
  Proceedings of the National Academy of Sciences}, p.~201818668, June 2019.

\bibitem{maurer_local_1998}
J.~Maurer and P.~Tabeling, ``Local investigation of superfluid turbulence,''
  {\em EPL (Europhysics Letters)}, vol.~43, pp.~29--34, July 1998.

\bibitem{salort_turbulent_2010}
J.~Salort, C.~Baudet, B.~Castaing, B.~Chabaud, F.~Daviaud, T.~Didelot,
  P.~Diribarne, B.~Dubrulle, Y.~Gagne, F.~Gauthier, A.~Girard, B.~H{\'e}bral,
  B.~Rousset, P.~Thibault, and P.-E. Roche, ``Turbulent velocity spectra in
  superfluid flows,'' {\em Physics of Fluids}, vol.~22, p.~125102, Dec. 2010.

\bibitem{rusaouen_detection_2017}
E.~Rusaouen, B.~Rousset, and P.-E. Roche, ``Detection of vortex coherent
  structures in superfluid turbulence,'' {\em EPL (Europhysics Letters)},
  vol.~118, no.~1, p.~14005, 2017.

\bibitem{fisher_andreev_2014}
S.~N. Fisher, M.~J. Jackson, Y.~A. Sergeev, and V.~Tsepelin, ``Andreev
  reflection, a tool to investigate vortex dynamics and quantum turbulence in
  3he-{B},'' {\em Proceedings of the National Academy of Sciences}, vol.~111,
  pp.~4659--4666, Mar. 2014.

\bibitem{guo_visualization_2014}
W.~Guo, M.~La~Mantia, D.~P. Lathrop, and S.~W. Van~Sciver, ``Visualization of
  two-fluid flows of superfluid helium-4,'' {\em Proceedings of the National
  Academy of Sciences}, vol.~111, pp.~4653--4658, Mar. 2014.

\bibitem{hall_rotation_1956}
H.~E. Hall and W.~F. Vinen, ``The {Rotation} of {Liquid} {Helium} {II}. {II}.
  {The} {Theory} of {Mutual} {Friction} in {Uniformly} {Rotating} {Helium}
  {II},'' {\em Proceedings of the Royal Society of London. Series A.
  Mathematical and Physical Sciences}, vol.~238, pp.~215--234, Dec. 1956.

\bibitem{bekarevich_phenomenological_1961}
I.~Bekarevich and I.~Khalatnikov, ``Phenomenological derivation of the
  equations of vortex motion in {He} {II},'' {\em Journal of Experimental and
  Theoretical Physics}, vol.~13, pp.~643--646, 1961.

\bibitem{lvov_energy_2006}
V.~S. L{\textquoteright}vov, S.~V. Nazarenko, and L.~Skrbek, ``Energy {Spectra}
  of {Developed} {Turbulence} in {Helium} {Superfluids},'' {\em Journal of Low
  Temperature Physics}, vol.~145, pp.~125--142, Nov. 2006.

\bibitem{schwarz_three-dimensional_1985}
K.~W. Schwarz, ``Three-dimensional vortex dynamics in superfluid {He}4:
  {Line}-line and line-boundary interactions,'' {\em Physical Review B},
  vol.~31, pp.~5782--5804, May 1985.

\bibitem{schwarz_three-dimensional_1988}
K.~W. Schwarz, ``Three-dimensional vortex dynamics in superfluid {He}4:
  {Homogeneous} superfluid turbulence,'' {\em Physical Review B}, vol.~38,
  pp.~2398--2417, Aug. 1988.

\bibitem{finne_intrinsic_2003}
A.~P. Finne, T.~Araki, R.~Blaauwgeers, V.~B. Eltsov, N.~B. Kopnin, M.~Krusius,
  L.~Skrbek, M.~Tsubota, and G.~E. Volovik, ``An intrinsic velocity-independent
  criterion for superfluid turbulence,'' {\em Nature}, vol.~424,
  pp.~1022--1025, Aug. 2003.

\bibitem{saffman_vortex_1992}
P.~G. Saffman, {\em Vortex {Dynamics}}.
\newblock Cambridge University Press, 1992.

\bibitem{baggaley_tree_2012}
A.~W. Baggaley and C.~F. Barenghi, ``Tree {Method} for {Quantum} {Vortex}
  {Dynamics},'' {\em Journal of Low Temperature Physics}, vol.~166, pp.~3--20,
  Jan. 2012.

\bibitem{baggaley_sensitivity_2012}
A.~W. Baggaley, ``The {Sensitivity} of the {Vortex} {Filament} {Method} to
  {Different} {Reconnection} {Models},'' {\em Journal of Low Temperature
  Physics}, vol.~168, pp.~18--30, July 2012.

\bibitem{sherwin-robson_local_2015}
L.~K. Sherwin-Robson, C.~F. Barenghi, and A.~W. Baggaley, ``Local and nonlocal
  dynamics in superfluid turbulence,'' {\em Physical Review B}, vol.~91,
  p.~104517, Mar. 2015.

\bibitem{morris_vortex_2008}
K.~Morris, J.~Koplik, and D.~W.~I. Rouson, ``Vortex {Locking} in {Direct}
  {Numerical} {Simulations} of {Quantum} {Turbulence},'' {\em Physical Review
  Letters}, vol.~101, p.~015301, July 2008.

\bibitem{baggaley_vortex-density_2012}
A.~W. Baggaley, J.~Laurie, and C.~F. Barenghi, ``Vortex-{Density}
  {Fluctuations}, {Energy} {Spectra}, and {Vortical} {Regions} in {Superfluid}
  {Turbulence},'' {\em Physical Review Letters}, vol.~109, p.~205304, Nov.
  2012.

\bibitem{monaghan_smoothed_1992}
J.~J. Monaghan, ``Smoothed {Particle} {Hydrodynamics},'' {\em Annual Review of
  Astronomy and Astrophysics}, vol.~30, pp.~543--574, Sept. 1992.

\bibitem{walmsley_quantum_2008}
P.~M. Walmsley and A.~I. Golov, ``Quantum and {Quasiclassical} {Types} of
  {Superfluid} {Turbulence},'' {\em Physical Review Letters}, vol.~100,
  p.~245301, June 2008.

\bibitem{babuin_quantum_2012}
S.~Babuin, M.~Stammeier, E.~Varga, M.~Rotter, and L.~Skrbek, ``Quantum
  turbulence of bellows-driven 4he superflow: {Steady} state,'' {\em Physical
  Review B}, vol.~86, p.~134515, Oct. 2012.

\bibitem{nore_decaying_1997}
C.~Nore, M.~Abid, and M.~E. Brachet, ``Decaying {Kolmogorov} turbulence in a
  model of superflow,'' {\em Physics of Fluids (1994-present)}, vol.~9,
  pp.~2644--2669, Sept. 1997.

\bibitem{araki_energy_2002}
T.~Araki, M.~Tsubota, and S.~K. Nemirovskii, ``Energy {Spectrum} of
  {Superfluid} {Turbulence} with {No} {Normal}-{Fluid} {Component},'' {\em
  Physical Review Letters}, vol.~89, p.~145301, Sept. 2002.

\bibitem{walmsley_dynamics_2014}
P.~Walmsley, D.~Zmeev, F.~Pakpour, and A.~Golov, ``Dynamics of quantum
  turbulence of different spectra,'' {\em Proceedings of the National Academy
  of Sciences}, vol.~111, pp.~4691--4698, Mar. 2014.

\end{thebibliography}
\end{document}